\documentclass{PoS}

\def\simge{\mathrel{%
       \rlap{\raise 0.511ex \hbox{$>$}}{\lower 0.511ex \hbox{$\sim$}}}}
\def\simle{\mathrel{
       \rlap{\raise 0.511ex \hbox{$<$}}{\lower 0.511ex \hbox{$\sim$}}}}

\title{Recent progress in lattice QCD at finite density }

\ShortTitle{Recent progress in lattice QCD at finite density }

\author{\speaker{Shinji Ejiri}%
\\
Physics Department, Brookhaven National Laboratory,
Upton, New York 11973, USA\\
        E-mail: \email{ejiri@quark.phy.bnl.gov}}

\abstract{
We review recent progress in lattice QCD at finite density. The phase diagram of QCD and the equation of state at finite temperature and density are discussed. In particular, we focus on the critical point terminating a first order phase transition line in the high density region. The critical point is one of the most interesting features that may be discovered in heavy-ion collision experiments. We summarize the current discussion on the existence of a critical point in the QCD phase diagram and discuss some attempts to find the critical point by numerical simulations. 
}

\FullConference{The XXVI International Symposium on Lattice Field Theory \\
		July 14 - 19, 2008\\
		Williamsburg, Virginia, USA}

\begin{document}

\section{Introduction}
\label{sec:intro} 

The primordial state of matter called quark-gluon plasma (QGP) is expected to be realized in extremely hot and dense mediums, and a lot of experimental efforts have been made to produce such a state in heavy-ion collision experiments \cite{YHM}. 
To understand QGP, theoretical studies by the first principle calculations of QCD at high temperature and density are important. 
At present, the lattice QCD simulation is the only systematic method to do so.
Many important properties of finite temperature QCD have been studied by the Monte-Carlo simulations \cite{DeTar08}. The studies at finite density had been known to be difficult until recently. 
However, recent technical developments allow us to extract information on those in the low density region.
In this report, we would like to review recent progress in lattice QCD at finite density. 
One of the most interesting studies is to investigate the phase structure of QCD at non-zero temperature and density. 
The QCD phase transition has been found to be crossover at zero density by simulations with staggered type quark actions \cite{Aoki06}. 
We expect that the nature of the phase transition will change to be of first order in the high density region, and it is very important to find the critical point terminating a first order phase transition line, since the critical point is one of the most interesting features that may be discovered in heavy-ion collision experiments.
The study of the equation of state (EoS) is also important. 
The numerical studies by lattice QCD simulations will be able to provide basic input for hydrodynamic calculations of the expansion of hot and dense matter generated in heavy-ion collisions. 

However, lattice QCD at non-zero density is known to have a serious problem. 
In a Monte Carlo simulation, we generate configurations of link variables 
$\{ U_{\mu}(x) \}$ with the probability in proportion to the weight factor 
$(\det M)^{N_{\rm f}} e^{-S_g}$ and the state density of $\{ U_{\mu}(x) \}$. 
Here, $M$ is the quark matrix and $S_g$ is the gauge action.
The expectation value of an operator ${\cal O}[U_{\mu}]$ is then evaluated 
by taking an average of ${\cal O}[U_{\mu}]$ over the generated 
configurations $\{ U_{\mu}(x) \}$.
\begin{eqnarray}
\langle {\cal O} \rangle_{(\beta)} 
\approx \frac{1}{N_{\rm conf.}} \sum_{ \{ U_{\mu}(x) \} } {\cal O}[U_{\mu}].
\end{eqnarray}
The quark matrix at zero density have the $\gamma_5$ Hermiticity 
and the Hermiticity guarantees that the quark determinant is real. 
However, the relation of the $\gamma_5$ Hermiticity changes to 
\begin{eqnarray}
M^{\dagger}(\mu_q) = \gamma_5 M(-\mu_q) \gamma_5. 
\label{eq:gamma5conj}
\end{eqnarray}
at finite quark chemical potential $(\mu_q)$. 
Then, the quark determinant becomes complex except for $\mu_q=0$; 
$(\det M(\mu_q))^* = \det M (-\mu_q) \neq \det M (\mu_q)$. 
Because the Boltzmann weight must be real and positive in the Monte-Carlo method, we cannot perform a simulation at finite density directly  
A popular method to deal with QCD at finite $\mu_q$ is the reweighting method. However, we will encounter another problem called the ``sign problem'' in the calculation at large $\mu_q$. 
The key point in the study of finite density QCD is to avoid this problem.

A lot of progresses have been obtained in this field. 
The equation of state in the low density region was studied in \cite{MILC07,MILC08,RBCB08,WHOT08}.
The sign problem is still one of the most important issues in the study of finite density lattice QCD. 
The nature of the sign problem was discussed in the random matrix model and the chiral perturbation theory \cite{spli06,spli07,Ste08}. 
Some trials to find the critical point at finite density were examined \cite{Krat05, eji07,eji08,Li08}. 
Moreover, a new algorithm based on stochastic quantization was proposed in \cite{Aarts08}. 
The phase structure in the high density region was studied in the strong coupling limit \cite{Miura08,From08}.
The hadronic fluctuations in the high temperature phase was also studied using an effective theory \cite{Hiet08}.
Moreover, the equation of state by chiral fermion actions was discussed in the high temperature limit \cite{Gav08,Hegde08}.
The phase structure of two-color QCD, which is free from the sign problem, has been studied. (See e.g. \cite{LomQM08,Hands07,Naka03} for a review.)
Among these topics, we want to focus on the equation of state and the critical point in the $(T, \mu_q)$ plane in this report. 
We discuss the equation of state and hadronic fluctuations in Sec.~\ref{sec:eos}. 
Some attempts to find the critical point are discussed in Sec.~\ref{sec:cpt}. 
A summary is given in Sec.~\ref{sec:summary}.

\section{Equation of state at finite density}
\label{sec:eos} 

In order to extract unambiguous signals for the QCD phase transition 
from the heavy-ion collisions, quantitative calculations 
from the first principles of QCD are indispensable. 
In particular, studies of the equation of state (EoS) can provide basic input for the analysis of the experimental data. 
Many studies have been done at finite temperature $(T)$ and zero chemical potentials $(\mu_q)$ \cite{DeTar08}. 
Also, recent developments of computational techniques enabled us to extend the study to small $\mu_q$.

Several years ago, systematic simulations for the study of the EoS at finite density have been performed by the Bielefeld-Swansea Collaboration using p4-imploved staggered quark action with rather heavy quark masses \cite{BS02,BS03,BS05}.
They found that the Taylor expansion method is useful for the EoS study in the low density region which is important for heavy-ion collisions. 
Moreover, they found large fluctuations in the quark number density at finite density. 
The temperature dependence of the quark number susceptibility $\chi_q$, which corresponds to the fluctuation of the quark number, changes qualitatively when $\mu_q$ becomes non-zero. For $\mu_q=0$ the susceptibility $\chi_q/T^2$ changes rapidly at the transition temperature but continues to increase monotonically. However, for $\mu_q \neq 0$ the quark number susceptibility develops a pronounced peak at the transition temperature. Such a behavior suggests the existence of a critical point in the $(T, \mu_q)$ phase diagram.

In this year, remarkable results were obtained by simulations near the physical quark mass point with improved staggered quark actions \cite{MILC07,MILC08,RBCB08}. 
The MILC Collaboration and the RBC-Bielefeld Collaboration studied the isentropic equation of state, i.e. the EoS along trajectories of constant entropy per baryon number.
There are also progresses in the study of fluctuations at finite density.
The RBC-Bielefeld Collaboration found that the enhancement of the quark number susceptibility becomes larger as the quark mass decreased \cite{RBCB08}. 
Moreover, the WHOT-QCD Collaboration performed simulations with a Wilson type quark action and studied the EoS at finite density \cite{WHOT08}.
They calculated the quark number susceptibility and confirmed the large fluctuation at $\mu_q \neq 0$.

\subsection{Taylor expansion method}
\label{sec:taylor}

The main problem in the study of QCD at finite density is that 
the Boltzmann weight is complex for $\mu_q \ne 0$. 
Because the Boltzmann weight must be real and positive if we want to generate configurations with the weight, the conventional Monte-Carlo method is not applicable at $\mu_q \ne 0$. 
One of the possible approaches to study the finite density QCD is performing a Taylor expansion of physical quantities in terms of $\mu_q$ around $\mu_q=0$ and calculating the expansion coefficients by numerical simulations at $\mu_q=0$ \cite{BS02,BS03,BS05,Miya02,GG1}.
Because the simulations at $\mu_q=0$ is free from the complex weight problem, 
the expansion coefficients, i.e. derivatives of physical quantities with respect to $\mu_q/T$, can be evaluated by a conventional Monte-Carlo simulation.
The pressure $(p)$ is obtained from the partition function $({\cal Z})$,
\begin{eqnarray}
\frac{p}{T^4} = \frac{1}{VT^3} \ln {\cal Z} \equiv \Omega, 
\label{eq:pdef}
\end{eqnarray}
and the calculations of the derivatives the partition function are basic measurements in the QCD thermodynamics, since most of thermodynamic quantities are given by the derivatives of $\Omega$.

We define the Taylor expansion coefficients as
\begin{equation}
\frac{p}{T^4} =
\sum_{i,j,k=0}^\infty c_{i,j,k}^{u,d,s}(T) \left(\frac{\mu_u}{T}\right)^i
\left(\frac{\mu_d}{T}\right)^j \left(\frac{\mu_s}{T}\right)^k,
\hspace{2mm}
c_{i,j,k}^{u,d,s} = 
\frac{1}{i!j!k!} \left.
\frac{\partial^{i+j+k} \Omega}{\partial(\mu_u/T)^i
\partial(\mu_d/T)^j \partial(\mu_s/T)^k} \right|_{\mu_{u,d,s}=0}.
\label{eq:cn}
\end{equation}
Here, $\mu_{u,d,s}$ are the chemical potentials for the u,d,s quarks, 
and $c_{0,0,0}^{u,d,s}(T)$ is the pressure at $\mu_u=\mu_d=\mu_s=0$. 
The coefficient $c_{i,j,k}^{u,d,s}(T)$ are computed by performing a simulation at $\mu_q=0$. The explicit forms of the Taylor expansion coefficients are given in \cite{MILC07,BS05}. 
We expect that QCD in the high temperature limit is described as free gas 
of quark and gluon 
and the $\mu_q$-dependence of $p/T^4$ is given only through terms 
of $\mu_q^2$ and $\mu_q^4$ for the free gas.
Therefore, the Taylor expansion may converge well in the high temperature region.

For the calculation of pressure at $\mu_q=0$, the integral method is commonly used. Using the thermodynamic relation Eq.~(\ref{eq:pdef}), the pressure is computed as 
\begin{equation}
p = \frac{T}{V} \int^{\beta}_{\beta_0} \! d\beta \, \frac{1}{\cal Z}
\frac{\partial {\cal Z}}{\partial \beta} 
= -\frac{T}{V} \int^{\beta}_{\beta_0} \! d\beta \left\langle
\frac{\partial S_{\rm lat}}{\partial \beta} \right\rangle . 
\end{equation}
Here, $S_{\rm lat}$ is the lattice action and $\langle \cdots \rangle$ 
is the thermal average with zero temperature contribution subtracted 
for the normalization of $p$. 
In multi-parameter cases such as full QCD, $\beta$ should be  
generalized to the position vector in the coupling parameter space. 
The initial point of integration $\beta_0$ is chosen in the low
temperature phase from the condition $p(\beta_0) \approx 0$.
The derivatives of $S_{\rm lat}$ with respect to $\beta$ and the quark 
mass are basically given by the Wilson loops and chiral condensate.

The energy density is obtained from the following equation, 
\begin{eqnarray}
\frac{\varepsilon - 3p}{T^4}
= \frac{1}{VT^2} \left. \frac{\partial \ln {\cal Z}}{\partial T} \right|_{\mu_q/T}
= \frac{N_t^3}{N_s^3} \left\langle a \left. 
\frac{\partial S_{\rm lat}}{\partial a} \right|_{\mu_q/T} \right\rangle, 
\hspace{3mm}
\left\langle a \frac{\partial S_{\rm lat}}{\partial a} \right\rangle 
=a \frac{\partial \beta}{\partial a}
\left\langle \frac{\partial S_{\rm lat}}{\partial \beta} \right\rangle, 
\label{eq:dsda}
\end{eqnarray}
where $a$ is the lattice spacing, the lattice size is $N_s^3 \times N_t$,
and $\beta$ is the position vector in the coupling parameter space for full QCD, again.
The density effect of $(\varepsilon - 3p)/T^4$ can be estimated by a Taylor expansion. The coefficients are given by the derivatives of Eq.~(\ref{eq:dsda}) with respect to $\mu_q/T$. 
The quark number density $n_{u,d,s}$ is calculated by
\begin{eqnarray}
\frac{n_{u,d,s}}{T^3} = 
\frac{1}{VT^3} \frac{\partial \ln {\cal Z}}{\partial (\mu_{u,d,s}/T)} 
= \frac{\partial (p/T^4)}{\partial (\mu_{u,d,s}/T)} 
\label{eq:qnd}
\end{eqnarray}
and Eq.~(\ref{eq:cn}).
We define the light quark number density as $n_q=n_u+n_d$. 
The susceptibilities of light Quark number $(\chi_q)$ and strange quark number $(\chi_s)$ are given by
\begin{eqnarray}
\frac{\chi_q}{T^2} 
= \left( \frac{\partial}{\partial (\mu_u/T)} 
+ \frac{\partial}{\partial (\mu_d/T)} \right) 
\frac{n_u + n_d}{T^3}, \hspace{5mm} 
\frac{\chi_s}{T^2} 
= \frac{\partial (n_s/T^3)}{\partial (\mu_s/T)}.
\end{eqnarray}
These susceptibilities correspond to the fluctuations of the quark numbers.
Moreover, the entropy density $s$ is given by the thermodynamic relation,
\begin{eqnarray}
\frac{s}{T^3}=
\frac{\varepsilon +p - \sum_{f=u,d,s} \mu_f n_f}{T^4} .
\end{eqnarray}
The chiral condensate is defined by the derivative of 
$\ln {\cal Z}$ with respect to the quark mass.

\subsection{Isentropic equation of state}
\label{sec:isentro} 

\begin{figure}[t]
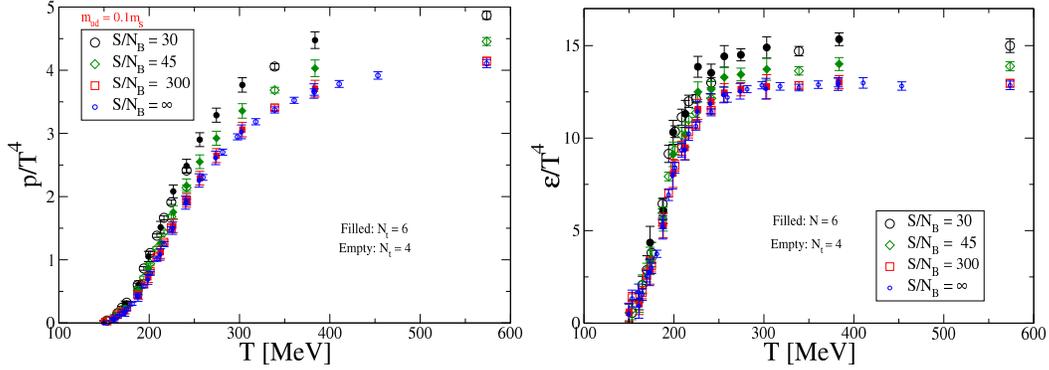

\begin{center}
\includegraphics[width=2.7in]{figs/PO6isentr.eps}
\includegraphics[width=2.7in]{figs/EO6Isentr.eps}
\vskip -0.2cm
\caption{Pressure (left) and energy density (right) vs. temperature along the lines of constant entropy per baryon number obtained by 2+1 flavor simulations with asqtad staggered fermion action \cite{MILC08}.
}
\label{fig:milc}
\end{center}
\vskip -0.3cm
\end{figure} 

\begin{figure}[t]
\begin{center}
\includegraphics[width=2.9in]{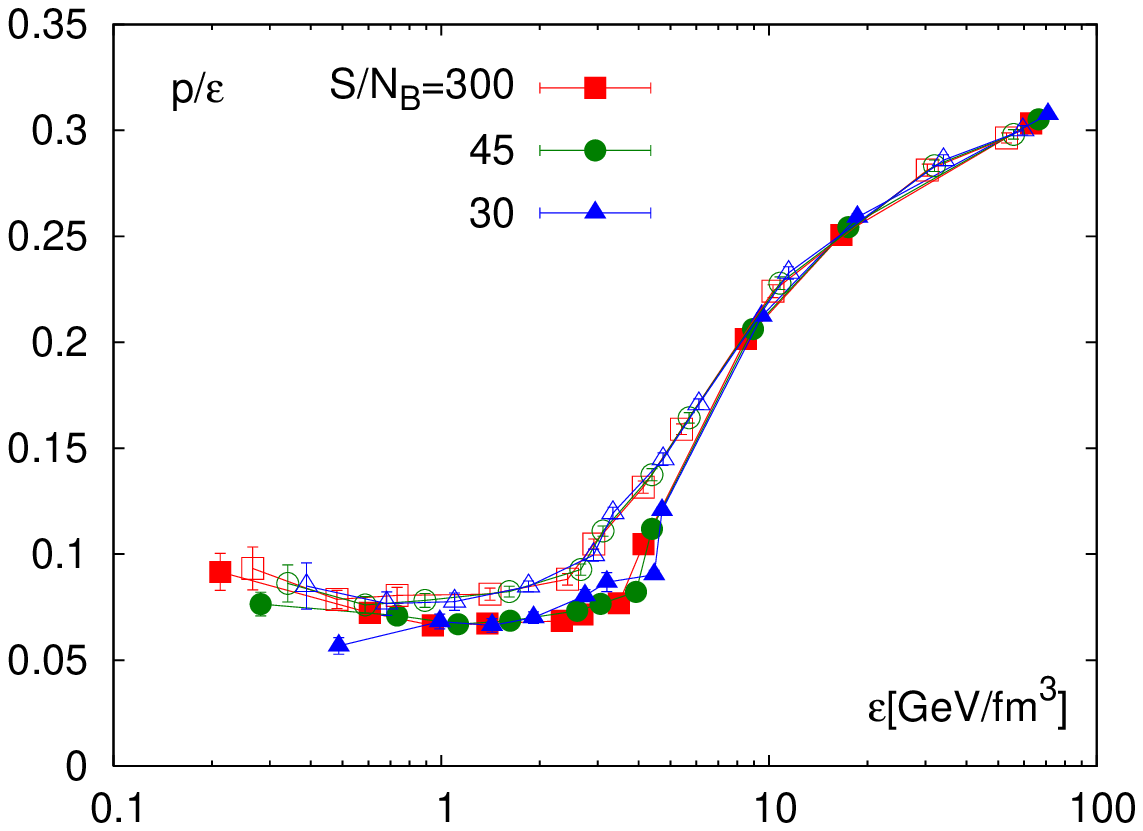}
\includegraphics[width=2.9in]{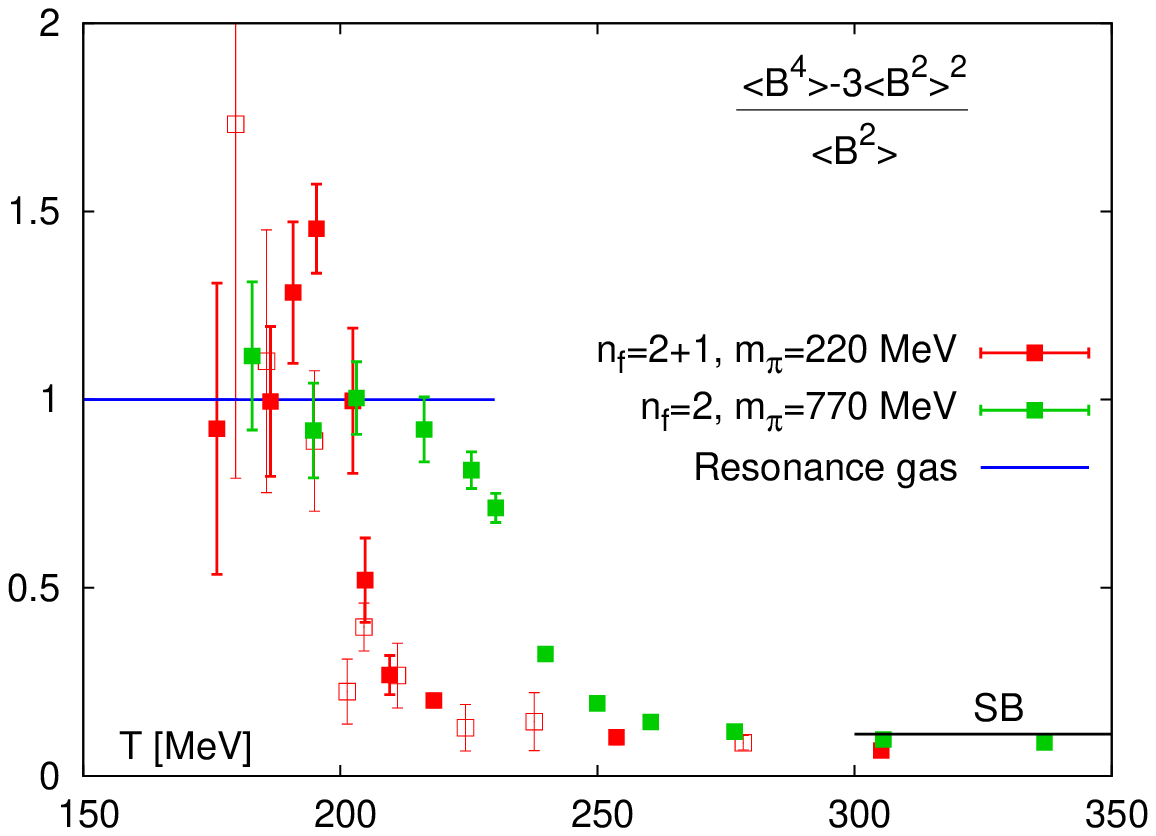}
\vskip -0.2cm
\caption{(left) The ratio of pressure and energy density as a function of energy density on isentropic trajectories obtained with p4fat3 staggered fermion action for $N_t=4$ (filled) and $6$ (open). 
(right) The ratio of $\chi_4^B$ and $\chi_2^B$ vs. temperature for $m_{\pi} \approx 220 {\rm MeV}$ \cite{RBCB08} and $m_{\pi} \approx 770 {\rm MeV}$ \cite{BS05}. }
\label{fig:rbcb}
\end{center}
\vskip -0.3cm
\end{figure} 

One of the most interesting results which have been obtained from heavy-ion collision experiments is that the experimental data is well-explained by a perfect fluid model without viscosity. This implies that a dense medium created in a heavy-ion collision expands without further generation of entropy after thermalization. 
Therefore, it is important to calculate the EoS with keeping the entropy $(S)$ per baryon number $(N_B)$ constant for the analysis of the experimental data \cite{isen06}. 

The MILC Collaboration and the RBC-Bielefeld Collaboration studied the isentropic equation of state by performing simulations near the physical quark mass point using improved staggered fermion actions. 
The isentropic expansion lines for matter created at RHIC, SPS and AGS energies correspond to $S/N_B \approx 300$, $S/N_B \approx 45$, and $S/N_B \approx 30$, respectively.
These values have been obtained by comparing experimental results for yields of various hadron species with hadron abundances in a resonance gas \cite{CleyRed}.
Measuring the Taylor expansion coefficients of the pressure, energy density, baryon number and entropy by Monte-Carlo simulations, they found the isentropic trajectories in the $(T, \mu_q, \mu_s)$ parameter space with an additional constraint $n_s=0$, where $\mu_u = \mu_d = \mu_q$.
They then calculated the energy density and pressure along the isentropic trajectory using the Taylor expansion coefficients.

The results of energy density and pressure computed by the MILC Collaboration are shown in Fig.~\ref{fig:milc} for each $S/N_B$ \cite{MILC07,MILC08}. 
The pion mass is about $m_{\pi} \approx 220 {\rm MeV}$, which is close to the physical pion mass. 
They used the asqtad quark action and successfully reduced the discretization error in the EoS. 
The filled and open symbols in these figures are the results on lattices with a temporal extent $N_t=6$ and $4$, respectively. The difference between them is found to be small.

The RBC-Bielefeld Collaboration also studied the isentropic equation of state performing 2+1 flavor simulations with $m_{\pi} \approx 220 {\rm MeV}$ using the p4fat3 action \cite{RBCB08}. The results of the pressure and energy density are consistent with the results by the MILC Collaboration. The right panel of Fig.~\ref{fig:rbcb} is the results of $p/\varepsilon$ plotted as a function of $\varepsilon$. 
The open symbols are the results from $N_t=6$ and the filled symbols are from $N_t=4$. 
They found that the density dependence of $p/\varepsilon$ is small when they plot as a function of $\varepsilon$. 
Also, the speed of sound in the dense medium, 
$c_s^2 = dp/d \varepsilon$, 
can be calculated by measuring the slope of $p/\varepsilon$ in this figure.

\subsection{Radius of convergence and hadronic fluctuations}
\label{sec:conver} 

Next, let us discuss the convergence radius of the Taylor series, 
$p/T^4 = \sum c_{i}^B (\mu_B/T)^i$, with 
$c_i^B=(1/3)(1/i!) (\sum_{f=u,d,s} \partial/\partial (\mu_f/T))^i 
\Omega|_{\mu_{u,d,s}=0}$.
We expect that the crossover transition at low density changes to a first order phase transition at a critical value of $\mu_B$. 
If there is such a critical point, the Taylor series does not converge at the critical point. The simplest way to estimate the radius of convergence $(\rho)$ is to calculate the ratio of the expansion coefficients.
We define 
\begin{eqnarray}
\rho = \lim_{n \to \infty} \rho_n, \hspace{5mm} 
\rho_n = \sqrt{|c_n^B/c_{n+2}^B|}.
\end{eqnarray}
In the region of $\mu_B/T < \rho$, $p/T^4$ is finite.
For the case of free quark gas expected in the high temperature limit, $c_n^B$ is zero for $n \geq 6$. A model described by resonances of hadron gas in the low temperature phase predicts $[p(\mu_B)-p(0)]/T^4 \propto \cosh(\mu_B/T)$ and  
$\rho_n=\sqrt{(n+2)(n+1)}$. 
Therefore, both the convergence radiuses in the high temperature and low temperature limits are infinity. On the other hand, by using an appropriate scaling ansatz for the free energy at $\mu_B=0$, one can show that $c_4$ will develop a cusp in the 2 flavor chiral limit with rather heavy strange quark mass. 
Hence, $c_4^B/c_2^B = \rho_2^{-2}$ should have a peak near the transition temperature and the radius of convergence may be short near $T_c$ when the u, d quark mass is sufficiently small. This implies that the distance to the critical values of $\mu_B/T$ may be estimated by measuring $\rho_n$ with rather small $n$.

The convergence radius $\rho_n$ have been studied in a 2 flavor simulation with a pion mass of $m_{\pi} \approx 770 {\rm MeV}$ \cite{BS05}. 
The results of $\chi_4^B/\chi_2^B = 12 c_4^B/c_2^B$ are shown by green symbols in Fig.\ref{fig:rbcb}, where $\chi_n^B \equiv n! c_n^B$. 
The result is consistent with the hadron resonance gas prediction at low temperature and with the free gas value at high temperature. 
However, there is no peak around $T_c$. Since $\chi_2^B$ also increases sharply just below $T_c$, $\chi_4^B/\chi_2^B$ does not increase near $T_c$ although $\chi_4^B$ itself has a pronounced peak. 
It is interesting to study the behavior of $\rho_n$ with small u, d quark masses near the physical point. The red symbols in Fig~\ref{fig:rbcb} are results of $\chi_4^B/\chi_2^B$ near the physical mass point in 2+1 flavor QCD obtained by the RBC-Bielefeld Collaboration \cite{RBCB08}. 
They observed a peak near $T_c$ beyond the hadron resonance gas value for $N_t=4$, which may be related to the critical point. 

Moreover, because $\chi_2^B = \chi_B/T^2$ and 
$\chi_4^B = \partial (\chi_B/T^2) / \partial (\mu_B/T)$ at $\mu_B=0$, the figure of $\chi_4^B/\chi_2^B$ indicates that the baryon number susceptibility $(\chi_B/T^2)$ increases more sharply near the transition point as the density is increased when u, d quark masses are small.

\subsection{Study by a Wilson type quark action}
\label{sec:wilson}

\begin{figure}[t]
\begin{center}
\includegraphics[width=2.4in]{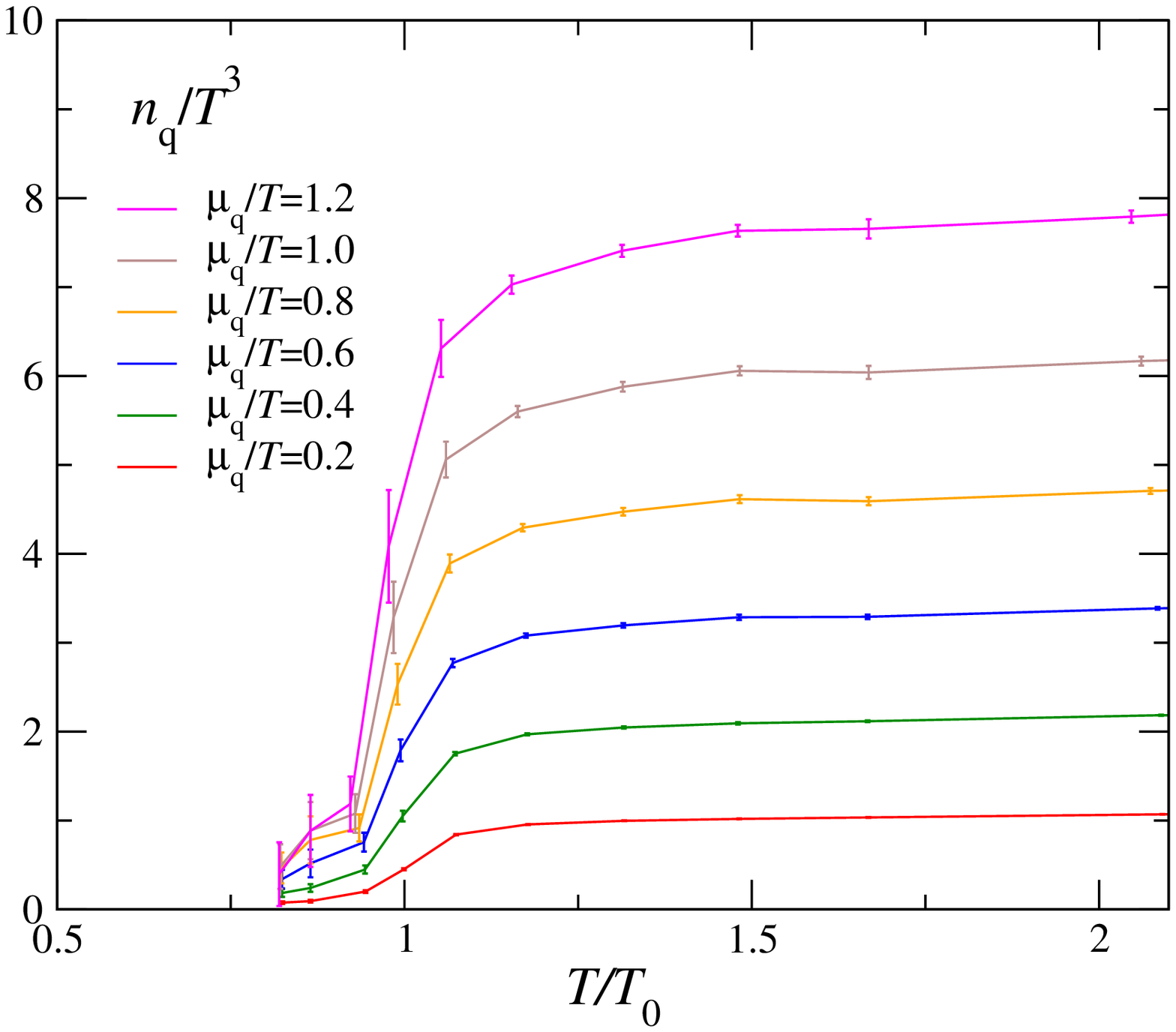}
\hskip 1.0cm
\includegraphics[width=2.4in]{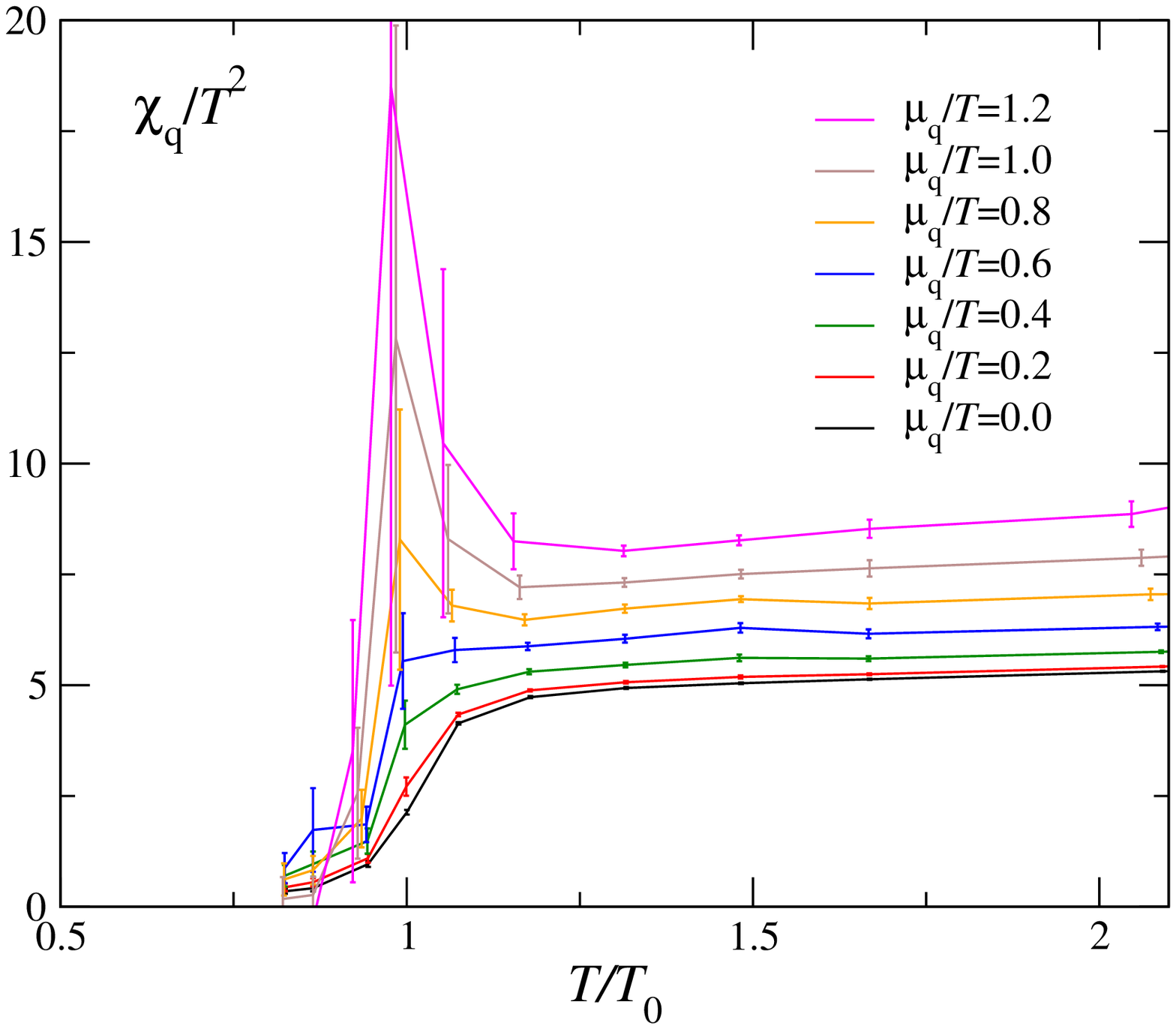}
\vskip -0.2cm
\caption{Quark number density (left) and quark number susceptibility (right) as functions of temperature and $\mu_q/T$ by a simulation with an improved Wilson quark action \cite{WHOT08}.
$T_0$ is $T_c$ at $\mu_q=0$.
}
\label{fig:whot}
\end{center}
\vskip -0.3cm
\end{figure} 

Most lattice QCD studies at finite temperature and density have been performed using staggered type quark actions with the fourth-root trick of the quark determinant. 
The theoretical base for the fourth-root trick is not confirmed.
Moreover, the scaling properties universal to the three-dimensional O(4) spin model, as expected from the effective sigma model, has not been confirmed in 2 flavor QCD. 
Therefore, it is important to carry out simulations adopting different lattice quark actions to control and estimate systematic errors due to lattice discretization. 

The WHOT-QCD Collaboration studied finite temperature and density QCD using the clover-improved Wilson quark action and the RG-improved Iwasaki gauge action. 
In contrast to the case of staggered quarks, the subtracted chiral condensate shows the scaling behavior with the critical exponents and scaling function of the O(4) spin model for this action \cite{cppacs1}, and the EoS at $\mu_q=0$ have been studied \cite{cppacs2}.
They performed simulations on $16^3 \times 4$ lattice along lines of constant physics with the mass ratio of pion and rho meson $m_{\pi}/m_{\rho}=0.65$ and $0.80$, and calculated the EoS at finite density \cite{WHOT08,WHOT07}.
Because the study by a Wilson quark action is more difficult than that by staggered quarks in general, some improvements are required.
They used a hybrid method of the Taylor expansion and the reweighting. 
Evaluating the quark determinant by the Taylor expansion up to $O(\mu_q^4)$, ${\cal Z}(\mu_q)/{\cal Z}(0)$ was computed. 
They then obtained the quark number density and the susceptibility by numerical differentiations with respect to $\mu_q$.
The results of the quark number density and the susceptibility are plotted in Fig.~\ref{fig:whot} as functions of $T$ for each $\mu_q/T$. 
These are quite similar to the results obtained by the previous staggered quark simulations. The quark number density increases sharply near $T_c$ and the slope becomes larger as $\mu_q/T$ increases. 
Also, they found that a peak in $\chi_q/T^2$ appears near $T_c$ for large $\mu_q/T$, suggesting the existence of the critical point.

\section{Critical point at finite density}
\label{sec:cpt}

In this section, we discuss the critical point terminating the first order phase transition line in the $(T, \mu_q)$ phase diagram sketched in Fig.~\ref{fig:massdep} (left).
The critical point is one of the most interesting features that may be discovered in heavy-ion collision experiments. 
We summarize the current discussion on the existence of the critical point in the QCD phase diagram. 

\subsection{Quark mass dependence of the critical point}
\label{sec:massdep}

\begin{figure}[t]
\begin{center}
\includegraphics[width=2.1in]{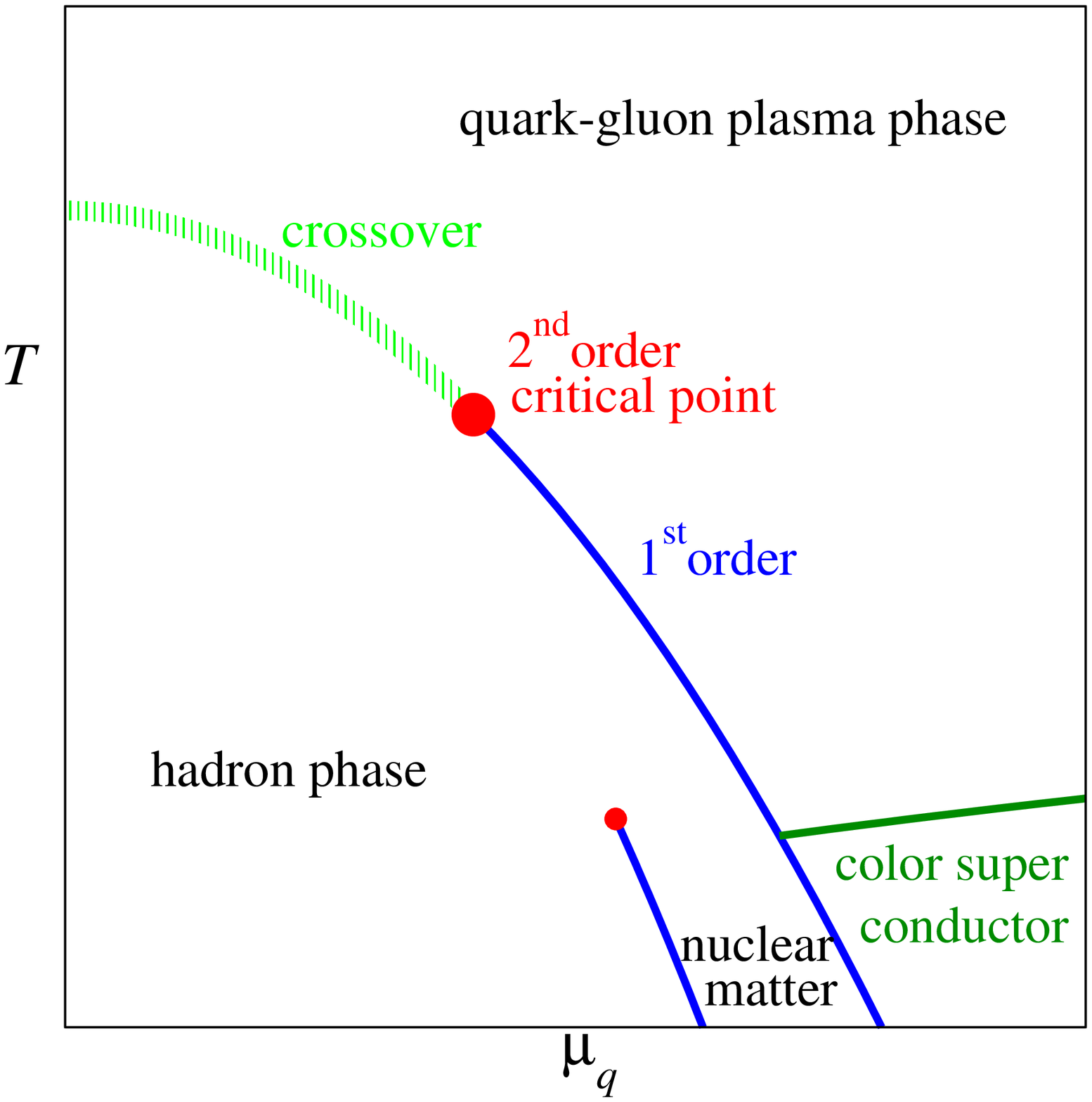}
\hskip 1.0cm
\includegraphics[width=2.2in]{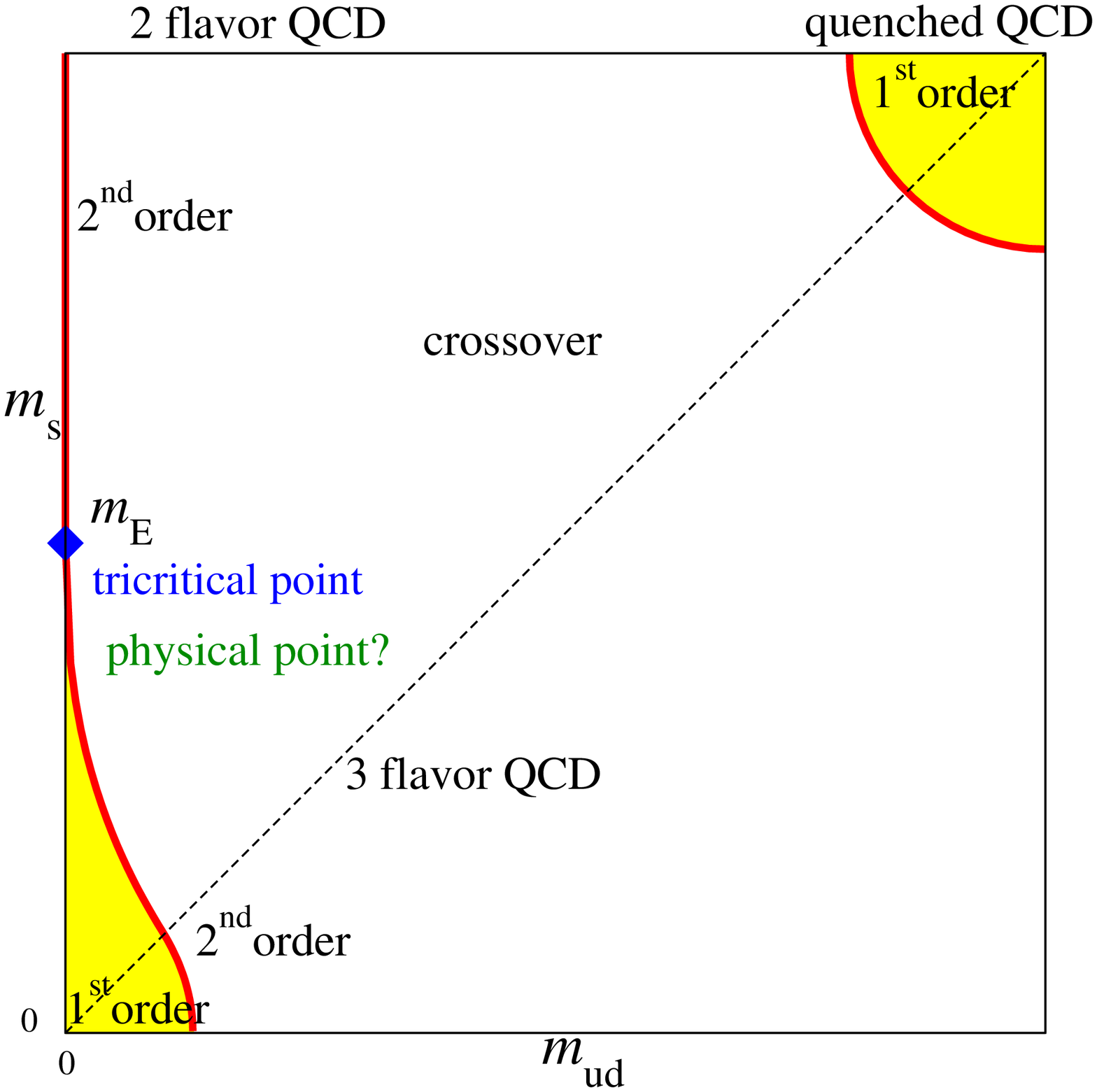}
\vskip -0.2cm
\caption{ Phase diagram in the $(T, \mu_q)$ plane (left) 
and quark mass dependence of the order of phase transitions, Columbia plot (right). 
}
\label{fig:massdep}
\end{center}
\vskip -0.3cm
\end{figure} 

\begin{figure}[t]
\begin{center}
\includegraphics[width=2.8in]{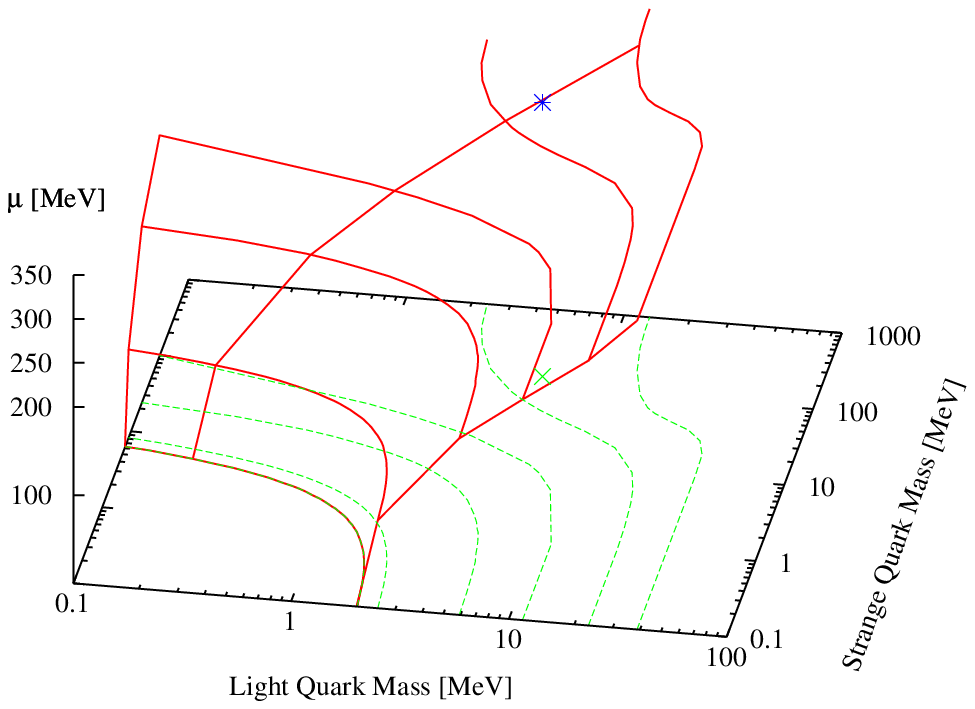}
\includegraphics[width=3.0in]{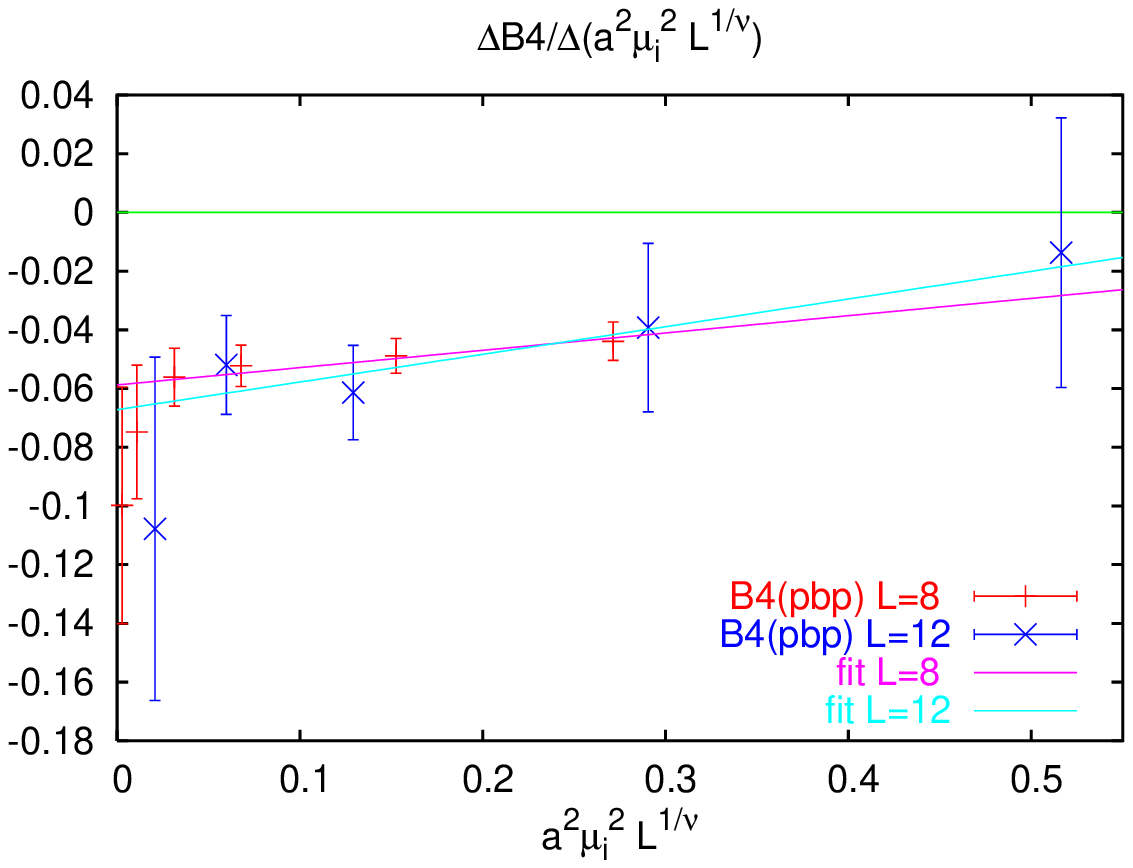}
\vskip -0.2cm
\caption{(left) Critical surface in the $(m_{ud}, m_s, \mu_q)$ parameter space predicted by the PNJL model \cite{Fuku08}. 
(right) $dB_4/d((\mu_i a)^2)$ vs. $(\mu_i a)^2$ obtained by a simulation with an imaginary chemical potential \cite{dFP3}. $L$ is the spatial extent $N_s$.
}
\label{fig:defo}
\end{center}
\vskip -0.3cm
\end{figure} 

The order of the phase transition depends on the quark mass for 2+1 flavor QCD. 
By changing the quark mass, the critical point at finite density can be shifted to the low density regime, where we can study it by a simulation. 
The expected nature of the phase transition at $\mu_q=0$ is summarized in the right panel of Fig.~\ref{fig:massdep}. The horizontal axis $m_{ud}$ is the u and d quark masses and the vertical axis $m_s$ is the strange quark mass. We expect that the phase transition of 2 flavor QCD in the chiral limit, $(m_{ud}, m_s)=(0, \infty)$, is of second order and that of 3 flavor QCD, $(m_{ud}, m_s)=(0,0)$, is of first order \cite{Pisarski}. 
The quenched limit, $(m_{ud}, m_s)=(\infty, \infty)$, is also of first order \cite{FOU,QCDPAX}.
The transition for 2 flavor QCD $(m_s= \infty)$ with finite $m_{ud}$ is crossover, and 2+1 flavor QCD has a second order critical line separating the first order region at small mass and the crossover region at large mass, which is shown by the bold red line in Fig.~\ref{fig:massdep} (right).

We can also discuss the nature of the phase transition at finite density. 
The left panel of Fig.~\ref{fig:defo} is a prediction of the critical surface in the $(m_{ud}, m_s, \mu_q)$ parameter space from the  Nambu-Jona-Lasinio model with the Polyakov loop (PNJL model) \cite{Fuku08}. 
The red lines indicate the critical surface which separates the first order and crossover regions.
We expect that the first order region becomes wider as $\mu_q$ increases and the crossover transition at low density changes to be of first order at high density for the physical quark masses.

\subsubsection*{Mean field argument near the tricritical point}

Let us start with a mean field analysis in the standard sigma model. 
We discuss the tricritical point on the $m_{ud}=0$ axis in Fig.~\ref{fig:massdep} (right), at which the second order critical line separates from the axis. 
In the vicinity of the tricritical point at $\mu_q=0$, the effective potential in terms of the chiral order parameter $\sigma$ is modeled by the following equation, 
\begin{eqnarray}
V_{\rm eff} (\sigma) = \frac{1}{2} a \sigma^2 + \frac{1}{4} b \sigma^4 
+ \frac{1}{6} c \sigma^6 -h \sigma, 
\end{eqnarray}
where we assume $c>0$ so that $V_{\rm eff}$ is bounded from below for large $|\sigma|$. 
The coefficients, $a, b$ and $h$ may be parameterized as 
\begin{eqnarray}
&& \! a= a_t t + a_{\mu} \mu_q^2, \hspace{3mm} 
b= b_s s + b_{\mu} \mu_q^2, \hspace{3mm} 
h= m_{ud} , \hspace{3mm}
t= \frac{T-T_c}{T_c}, \hspace{3mm} 
s= \frac{m_E -m_s}{m_E},
\label{eq:sigmapara}
\end{eqnarray}
where $m_E$ is $m_s$ at the tricritical point. 
The coefficient $b$ controls the order of phase transition. 
Assuming a symmetry under $\mu_q$ to $-\mu_q$, the leading contribution to $b$ must be $\mu_q^2$ at low density. 

Since the effective potential is $O(\sigma^4)$ on the second order critical surface, 
\begin{eqnarray}
\frac{\partial^n V_{\rm eff}}{\partial \sigma^n} = 0, \hspace{5mm}
(n=1,2,3).
\end{eqnarray}
Solving these equation, we obtain 
\begin{eqnarray}
\pm h= \frac{8c}{3} \left( \frac{a}{5c} \right)^{5/4}, \hspace{5mm} 
\pm h= \frac{8c}{3} \left( \frac{-3b}{10c} \right)^{5/2}, \hspace{5mm} 
(a \geq 0, b \leq 0).
\end{eqnarray}
The critical surface in the $(m_{ud}, m_s, \mu_q)$ space is described by 
\begin{eqnarray}
c_{ud} m_{ud}^{2/5} + c_s (m_E - m_s) + \mu_q^2 =0.
\label{eq:mcsur}
\end{eqnarray}
with appropriate constants $c_{ud}$ and $c_s$. 
The strange quark mass dependence and the $\mu_q$ dependence of the critical light quark mass $m_{ud}^c$ around the tricritical point are 
\begin{eqnarray}
m_{ud}^c \sim (m_E - m_s)^{5/2}, \hspace{5mm} 
m_{ud}^c \sim \mu_q^5.
\end{eqnarray}
The first equation describes the critical line on the $\mu_q=0$ plane sketched in Fig.~\ref{fig:massdep} (right). 
We expect from the first equation that the critical $m_{ud}$ increases very slowly as $m_s$ decreases. Similarly, the second equation suggests that the chemical potential dependence of the critical surface $m_{ud}^c (\mu_q)$ is also small in the low density region, since the $\mu_q$ dependence starts from a term of $\mu_q^5$ at $m_{ud}=0$. 
The information of the critical surface is important to know the order of phase transition for the real world, and the critical surface can be measured near the critical line at $\mu_q=0$ because the study by Monte-Carlo simulations is possible in the low density region.

\subsubsection*{Numerical study of the critical surface}

To investigate the critical surface, some groups performed simulations near the critical quark mass at $\mu_q=0$ in QCD with 3 flavors having degenerate quark masses, $m_{ud}=m_s$, and studied the $\mu_q$ dependence of the critical quark mass $m_c(\mu_q)$. 
For extrapolating $m_c(\mu_q)$, 
an approach on the basis of the Taylor expansion in terms of 
$\mu_q/T$ \cite{crtpt} and that of the imaginary chemical potential 
\cite{dFP2} have been developed.
Moreover, a study of phase-quenched finite density QCD, 
in which the effect from the complex phase of the quark determinant is neglected, has been discussed in \cite{KS07}.
As we expect form the mean field argument, the $\mu_q$ dependence of the critical mass have been found to be small in the low density region. 

Recently, an interesting result was obtained by de Forcrand and Philipsen \cite{dFP3}. 
They studied the $\mu_q$-dependence of the critical quark mass for QCD with 3 flavors of standard staggered quarks very precisely and found that the critical line moves towards lighter quark masses as a function of $\mu_q^2$.
They performed simulations with an imaginary chemical potential, $\mu=\mu_q a = i \mu_i a$, on an $N_s \times N_t = 8^3 \times 4$ and $12^3 \times 4$ lattices. 
Because $(\det M(\mu))^* = \det M (-\mu^*)$ for a complex $\mu$, the quark determinant is real if $\mu$ is purely imaginary, hence the simulations are possible. 
In order to identify $m_c (\mu_i)$, they computed the fourth order Binder cumulant constructed from the chiral condensate, 
$B_4= \langle (\delta \bar{\psi} \psi)^4 \rangle / 
\langle (\delta \bar{\psi} \psi)^2 \rangle^2$. 
It has been verified that the critical point belongs to 
the Ising universality class and the value of $B_4$ at $T_c$ for $m_c$ 
is the same with that of 3-dimensional Ising model, $B_4=1.604, $ \cite{KLS01}.
In the low density region, the analytic continuation from imaginary $\mu$ to real $\mu$ is performed assuming
\begin{eqnarray}
B_4=1.604 + b_{10} (m-m_c^0) + b_{01} \mu^2 + b_{02} \mu^4 + \cdots .
\label{eq:b4te}
\end{eqnarray}
Since $b_{10} (m-m_c^0) + b_{01} \mu^2 =0$ along the critical line in the leading order, one can estimate the curvature of the critical quark mass at $\mu=0$ by 
$\partial m_c / \partial (\mu^2) \approx -b_{01}/b_{10}$.
Moreover, because $B_4$ increases when the first order phase transition changes to crossover, $b_{10}$ is positive.
Hence, the curvature is positive (negative) if $b_{01}$ is negative (positive).
The results of $\partial B_4 / \partial ((\mu_i a)^2)$ is plotted as a function of $\mu_i^2$ in Fig.~\ref{fig:defo} (right). $b_{01}$ is given by 
$b_{01}=- \lim_{\mu_i^2 \to 0} \partial B_4 / \partial ((\mu_i a)^2)$. 
From this figure, $\partial m_c / \partial (\mu^2)$ is found to be negative. 
This result contradicts to the naive expectation.

In this situation, a simple extrapolation of the critical surface to the physical quark mass point is difficult because the first order region in the $(m_{ud}, m_s)$ plane becomes smaller as $\mu_q$ increases if we do not consider higher order terms of $\mu_q$ in the Taylor expansion of $m_c (\mu_q)$. 
To understand the critical surface in the $(m_{ud}, m_s, \mu_q)$ space, studies in a wide range of the chemical potential may be necessary.
The analytic continuation in the imaginary chemical potential approach is usually based on the Taylor expansion of $B_4$ and $m_c(\mu_q)$, e.g. Eq.~(\ref{eq:b4te}). 
One of the possible improvements to study in the wide range is to use another assumption which is based on a phenomenological model. 
The analytic continuation with various assumptions has been discussed in \cite{Lom07,Cea07}.

\subsection{Reweighting method and Sign problem}
\label{sec:rewei}

\begin{figure}[t]
\begin{center}
\includegraphics[width=2.5in]{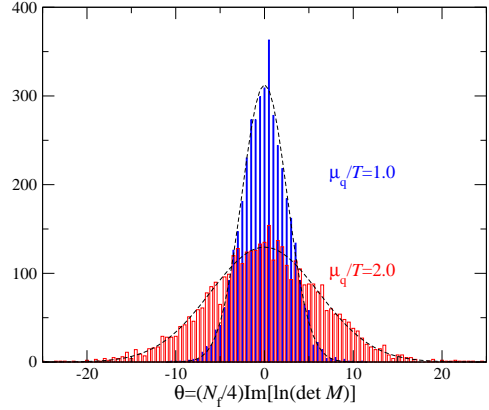}
\vskip -0.2cm
\caption{ Histograms of the complex phase $\theta$ \cite{eji07}. 
Dashed lines are the fit results by Gaussian functions.}
\label{fig:theta}
\end{center}
\vskip -0.3cm
\end{figure} 

Because the Boltzmann weight is complex at $\mu_q \neq 0$, the Monte-Carlo method is not applicable directly.
A popular approach to avoid this problem is the reweighting method 
\cite{Bar97,FK1}. 
We perform simulations at $\mu_q=0$, and incorporate the remaining
part of the correct Boltzmann weight for finite $\mu_q$ in the calculation
of expectation values. 
Expectation values $\langle {\cal O} \rangle$ at $(\beta, \mu_q)$ are thus
computed by a simulation at $(\beta, 0)$ using the following identity: 
\begin{eqnarray}
\langle {\cal O} \rangle_{(\beta, \mu_q)} 
= \frac{\left\langle {\cal O} 
\left( \det M(\mu_q) / \det M(0) \right)^{N_{\rm f}}
\right\rangle_{(\beta,0)}}{ \left\langle
\left( \det M(\mu_q) / \det M(0) \right)^{N_{\rm f}}
\right\rangle_{(\beta,0)}}. 
\label{eq:murew}
\end{eqnarray} 
This is the basic formula of the reweighting method. 
However, because $\det M(\mu_q)$ is complex, 
the calculations of the numerator and denominator in Eq.~(\ref{eq:murew}) 
becomes in practice increasingly more difficult for larger $\mu_q$. 
If the typical value of the complex phase of the quark determinant $\theta$ becomes larger than $\pi/2$, the real part of $e^{i \theta}$ $(=\cos \theta)$ changes its sign frequently.
Eventually both the numerator and denominator of Eq.~(\ref{eq:murew}) become smaller than their statistical errors and Eq.~(\ref{eq:murew}) can no longer be evaluated. We call it the ``sign problem''.

We define the phase of the quark determinant $\theta$ by 
the imaginary part of $N_{\rm f} \ln \det M(\mu_q)$ 
in the framework of the Taylor expansion. 
In this framework, $\ln \det M(\mu_q)$ can be separated into real and 
imaginary parts easily because the even derivatives of $\ln \det M(\mu_q)$ 
are real and the odd derivatives are purely imaginary \cite{BS02}. 
The complex phases $\theta$ are thus given by 
\begin{eqnarray}
\theta & = & N_{\rm f} {\rm Im} \left[ \ln (\det M) \right] 
= N_{\rm f} \sum_{n=0}^{\infty} \frac{1}{(2n+1)!} 
{\rm Im} \frac{\partial^{2n+1} (\ln \det M)}{\partial (\mu_q/T)^{2n+1}} 
\left( \frac{\mu_q}{T} \right)^{2n+1} , 
\label{eq:theta}
\end{eqnarray} 
where one must replace $N_{\rm f}$ in these equations to $N_{\rm f}/4$ 
when one uses a staggered type quark action. 
In Fig.~\ref{fig:theta}, we plot histograms of $\theta$ calculated at the pseudo-critical temperature $(\beta=3.65)$ for $\mu_q/T=1.0$ and $2.0$ using the data of the Taylor expansion coefficients up to $O(\mu_q^5)$ obtained with 2 flavors of p4-improved staggered quarks in \cite{BS05}.
The application range of the reweighting method can be estimated from the histogram. 
When the width of the distribution is larger than $O(\pi)$, the phase factor changes its sign frequently. Then, the sign problem happens and the reweighting method does not work. 
Here, it is worth noticing that $\theta$ corresponds to the complex 
phase of the quark determinant however 
this quantity is not restricted to the range from $-\pi$ to $\pi$ because 
there is no reason that the imaginary part of $\ln \det M$ 
in Eq.~(\ref{eq:theta}) must be in the finite range. 
An interesting point which is found from this figure is that these histograms seem to be almost Gaussian functions. 
We fit these data by Gaussian functions, $\sim \exp (-x \theta^2)$, 
with a fit parameter $x$. 
The dashed lines in Fig.~\ref{fig:theta} are the fit results. 
We find that the histogram of $\theta$ is well-represented 
by a Gaussian function. 
Such a Gaussian distribution is expected when the system size is sufficiently large in comparison to the correlation length due to the central limit theorem.
Moreover, the Gaussian distribution of $\theta$ has been discussed in chiral perturbation theory \cite{spli07}.

Once we assume a Gaussian distribution for $\theta$, 
the problem of complex weights can be avoided \cite{eji07}.
We calculate the expectation value of 
$\left\langle F e^{i\theta} \right\rangle_{(T, \mu_q=0)}$, 
where $F=|\det M(\mu_q)/ \det M(0)|^{N_{\rm f}}$ 
and ${\cal O} |\det M(\mu_q)/ \det M(0)|^{N_{\rm f}}$ 
for the denominator and numerator of Eq.~(\ref{eq:murew}), respectively. 
If the operator ${\cal O}$ is real, the complex phase is given by Eq.~(\ref{eq:theta}). 
We introduce the probability distribution $\bar{w}$ as 
a function of $F$ and $\theta$: 
\begin{eqnarray}
\bar{w}(F', \theta') \equiv 
\int {\cal D}U \delta (F'-F) \delta (\theta' - \theta) 
(\det M(0))^{N_{\rm f}} e^{-S_g},
\end{eqnarray}
where $\delta(x)$ is the delta function. 
The distribution function itself is defined as an expectation value 
at $\mu_q=0$, 
however $F$ and $\theta$ are functions of $\mu_q/T$ obtained by a simulation at $\mu_q=0$. 

Since the partition function is real even at non-zero density, 
the distribution function has the symmetry under 
the change from $\theta$ to $-\theta$.
Therefore, the distribution function is a function of $\theta^2$, 
e.g., $\bar{w}(\theta) \sim \exp[-(a_2 \theta^2 +a_4 \theta^4 
+a_6 \theta^6 + \cdots)].$
If the fluctuations of the phase is small at small $\mu_q$, the $\theta^2$ term gives a dominant contribution.
Moreover, as we discussed, the distribution function is well-approximated 
by a Gaussian function:
\begin{eqnarray}
\bar{w}(F, \theta) \approx \sqrt{\frac{a_2 (F)}{\pi}} 
\bar{w}'(F) \exp \left[-a_2 (F) \theta^{2} \right].
\label{eq:phap}
\end{eqnarray}
We assume this distribution function in terms of
$\theta$ when $F$ is fixed.
The coefficient $a_2 (F)$ is given by 
$1/(2a_2 (F')) =
\left\langle \theta^2 \delta (F'-F) \right\rangle_{(T, \mu_q=0)}/ \left\langle 
\delta (F'-F) \right\rangle_{(T, \mu_q=0)}
\equiv \left\langle \theta^2 \right\rangle_{F'}$.

Then, the integration over $\theta$ can be carried out easily,
\begin{eqnarray}
\left\langle F(\mu_q) e^{i \theta} \right\rangle_{(T, \mu_q=0)}
\approx \frac{1}{\cal Z} \int dF \ \bar{w}'(F) e^{-1/(4a_2)} F 
= \left\langle F(\mu_q) \ e^{- \left\langle \theta^2 \right\rangle_{F}/2} 
\right\rangle_{(T, \mu_q=0)}.
\label{eq:den12}
\end{eqnarray}
Since $\theta$ is roughly proportional to the size of the quark 
matrix $M$, the value of $\left\langle \theta^2 \right\rangle_{F}$
becomes larger as the volume increases as well as increasing $\mu_q$. 
Therefore, the phase factor in 
$\left\langle F(\mu_q) e^{i \theta} \right\rangle$ decreases 
exponentially as a function of the volume and $\mu_q$. 
However, the operator in Eq.~(\ref{eq:den12}) is always real and 
positive for each configuration, hence the expectation value of 
$\left\langle F(\mu_q) e^{i \theta} \right\rangle$
is always larger than its statistical error in this method.
Therefore, the sign problem is completely avoided if we can assume 
the Gaussian distribution of $\theta$.

\subsection{Plaquette effective potential}
\label{sec:poten}

One of the most primitive approaches to identify the order of the phase transition is to investigate the histogram of a typical quantity such as plaquette, Polyakov loop or chiral condensate by Monte-Caro simulations. 
For the case of the plaquette $(P)$, the distribution function, i.e. the histogram, is defined by
\begin{eqnarray}
w(P') = \int {\cal D} U \ \delta(P'-P) \ (\det M)^{N_{\rm f}} 
e^{6\beta N_{\rm site} P}. 
\label{eq:pdist}
\end{eqnarray}
For later discussions, we define the average plaquette $P$ as 
$ P \equiv -S_g/(6 \beta N_{\rm site})$, where 
$N_{\rm site} \equiv N_s^3 \times N_t$.
This is a linear combination of Wilson loops for an improved action.
If there is a first order phase transition point, where two different 
states coexist at the transition point, the histogram must have two 
peak at two different values of $P$ corresponding to the hot and cold phases.
Such studies have been done to confirm that the phase transition of SU(3) pure gauge theory is of first order, and the double peaked distribution have been obtained at the transition point \cite{FOU,QCDPAX}.
This method is equivalent to other methods to identify the order of phase transitions by the Binder cumulant and by the Lee-Yang zero \cite{eji05}. 

\begin{figure}[t]
\begin{center}
\includegraphics[width=2.0in]{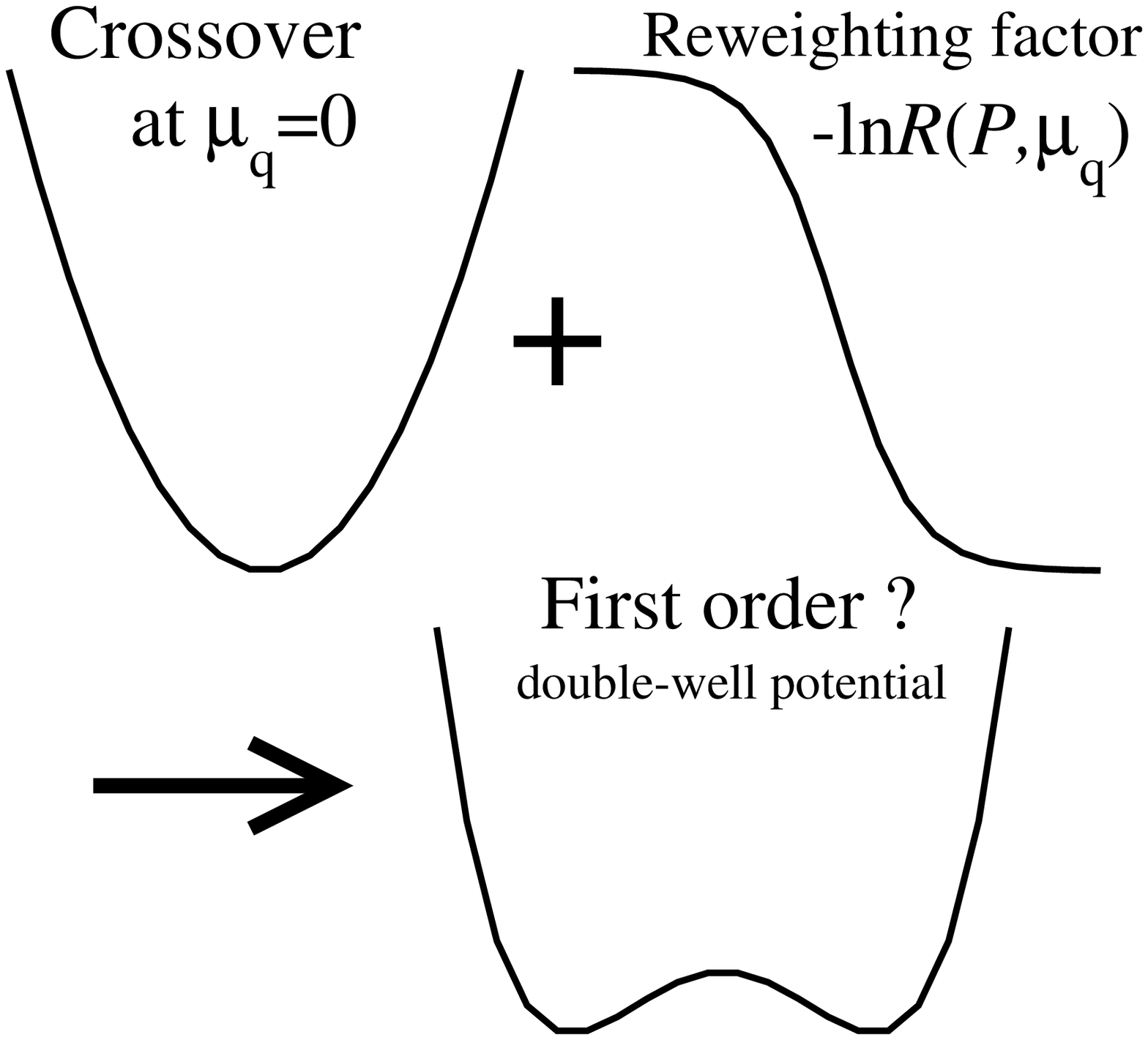}
\hskip 1.0cm
\includegraphics[width=2.7in]{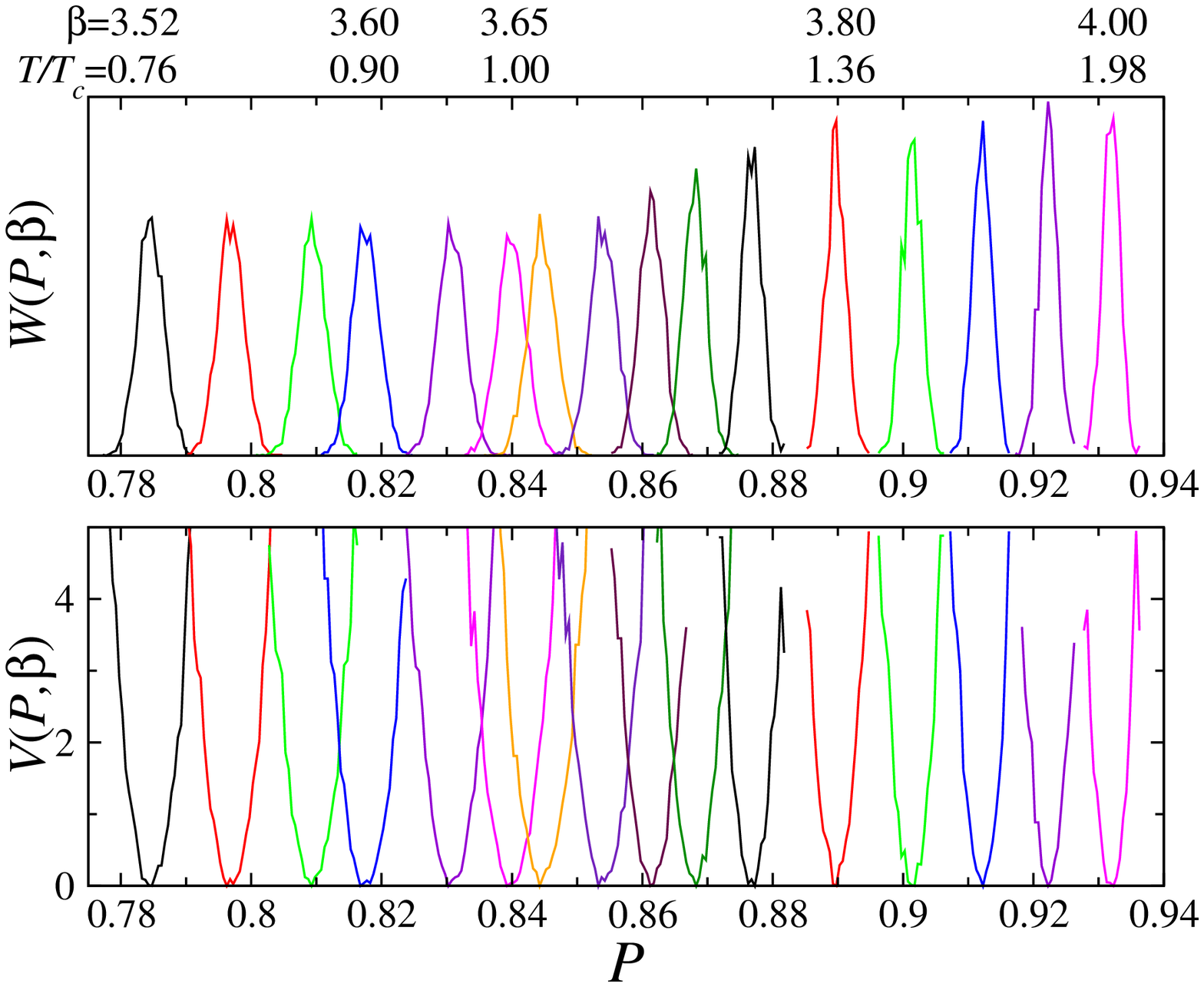}
\vskip -0.2cm
\caption{(left) Schematic illustration of the effective potential and the reweighting factor. 
(right) Plaquette histogram and the effective potential at $\mu_q=0$ \cite{eji07}.}
\label{fig:plaq}
\end{center}
\vskip -0.3cm
\end{figure}

The argument of the distribution was extended to the case of finite density QCD by \cite{eji07}.
Here and hereafter, we restrict ourselves to discuss only the case when 
the quark matrix does not depend on $\beta$ explicitly
for simplicity.
The partition function can be rewritten as 
\begin{eqnarray}
{\cal Z}(\beta, \mu_q) = \int R(P,\mu_q) w(P,\beta) \ dP,
\label{eq:rewmu}
\end{eqnarray}
where $w(P,\beta)$ is defined in Eq.~(\ref{eq:pdist}) at $\mu_q=0$ and 
$R(P,\mu_q)$ is the reweighting factor for finite $\mu_q$ 
defined by 
\begin{eqnarray}
R(P',\mu_q) \equiv 
\frac{\int {\cal D} U \ \delta(P'-P) (\det M(\mu_q))^{N_{\rm f}}}{
\int {\cal D} U \ \delta(P'-P) (\det M(0))^{N_{\rm f}}}
= \frac{ \left\langle \delta(P'-P) 
\left( \det M(\mu_q) / \det M(0) \right)^{N_{\rm f}}
\right\rangle_{(\beta, 0)} }{
\left\langle \delta(P'-P) \right\rangle_{(\beta, 0)}}.
\label{eq:rmudef}
\end{eqnarray}
This $R(P, \mu_q)$ is independent of $\beta$ and 
can be measured at any $\beta$. 
Here, $\left\langle \cdots \right\rangle_{(\beta, 0)}$ means 
the expectation value at $\mu_q=0$. In this method, 
all simulations are performed at $\mu_q=0$ and the effect of 
finite $\mu_q$ is introduced though the operator 
$\det M(\mu_q) / \det M(0)$ measured on the configurations 
generated by the simulations at $\mu_q=0$.
The distribution function for non-zero $\mu_q$ is 
$R(P, \mu_q) w(P,\beta)$, and thus the effective potential is defined by 
\begin{eqnarray}
V_{\rm eff}(P, \beta, \mu_q) \equiv -\ln [R(P, \mu_q) w(P, \beta)] 
= -\ln R(P, \mu_q) + V_{\rm eff}(P, \beta, 0).
\label{eq:potmu}
\end{eqnarray}
The shape of the effective potential can then be investigated 
at $\mu_q \neq 0$ once $R(P, \mu_q)$ is obtained.
A schematic illustration of $V_{\rm eff}(P)$ is shown in Fig.~\ref{fig:plaq} (left).

First, the peak position of the distribution function moves as $\mu_q$ changes,
which is determined by solving 
\begin{eqnarray}
\frac{\partial V_{\rm eff}}{\partial P} (P, \beta, \mu_q) 
= \frac{\partial V_{\rm eff}}{\partial P} (P, \beta, 0) 
- \frac{\partial (\ln R)}{\partial P} (P, \mu_q) =0.
\label{eq:minmuq}
\end{eqnarray} 
Then, the effect from $\mu_q$ to the peak position is the same as that when $\beta$ (temperature) is changed. 
From the definition at $\mu_q=0$, 
the weight $w(P, \beta)$ and the effective potential becomes 
\begin{eqnarray}
w(P, \beta_{\rm eff}) = e^{6 (\beta_{\rm eff} - \beta) N_{\rm site} P} w(P, \beta), \hspace{3mm} 
V_{\rm eff}(P,\beta_{\rm eff},0)
= V_{\rm eff}(P,\beta,0) -6 (\beta_{\rm eff} - \beta) N_{\rm site} P,
\label{eq:vrewbeta}
\end{eqnarray} 
under a change from $\beta$ to $\beta_{\rm eff}$. 
Hence, the change from $\beta$ to 
\begin{eqnarray}
\beta_{\rm eff}(\mu_q) 
\equiv \beta + (6N_{\rm site})^{-1} \partial (\ln R)/\partial P 
\label{eq:betaeff}
\end{eqnarray}
corresponds $(\beta, 0)$ to $(\beta, \mu_q)$ for the determination of the minimum of $V_{\rm eff} (P)$.
As we will see, the slope of $\ln R$ is positive. 
This explains why the phase transition happens when the density is 
increased as well as with increasing temperature $(\beta)$.

Moreover, we find from Eq.~(\ref{eq:vrewbeta}) that 
the curvature of $V_{\rm eff}(P)$ does not change under the change of $\beta$. 
This means that the curvature of $V_{\rm eff}(P)$ is independent of $\beta$. 
The critical value of $\mu_q$ can be estimated by measuring the curvature of 
the effective potential, 
\begin{eqnarray}
\frac{\partial^2 V_{\rm eff}}{\partial P^2} (P, \mu_q) 
= - \frac{\partial^2 (\ln R)}{\partial P^2} (P, \mu_q) 
+ \frac{\partial^2 V_{\rm eff}}{\partial P^2} (P, 0) =0. 
\end{eqnarray}
Because the curvature of $V_{\rm eff}(P, \beta, 0)$ at $\mu_q=0$ is positive 
and the curvature of $V_{\rm eff}(P, \beta, \mu_q)$ at a second order phase 
transition point is zero, 
the parameter range where $-\ln R(P, \mu_q)$ has negative 
curvature is required for the existence of the critical point.

\begin{figure}[t]
\begin{center}
\includegraphics[width=2.6in]{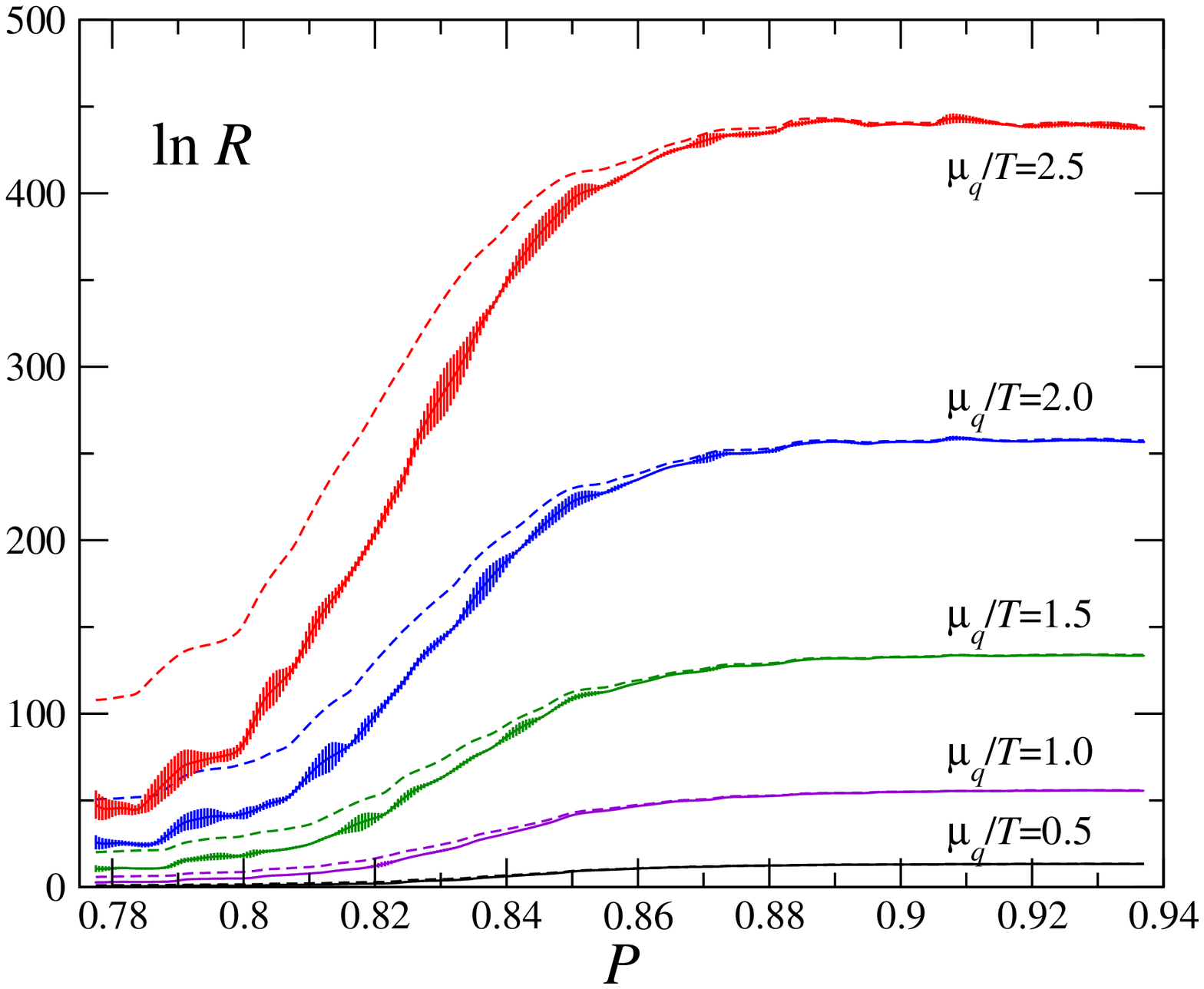}
\includegraphics[width=2.7in]{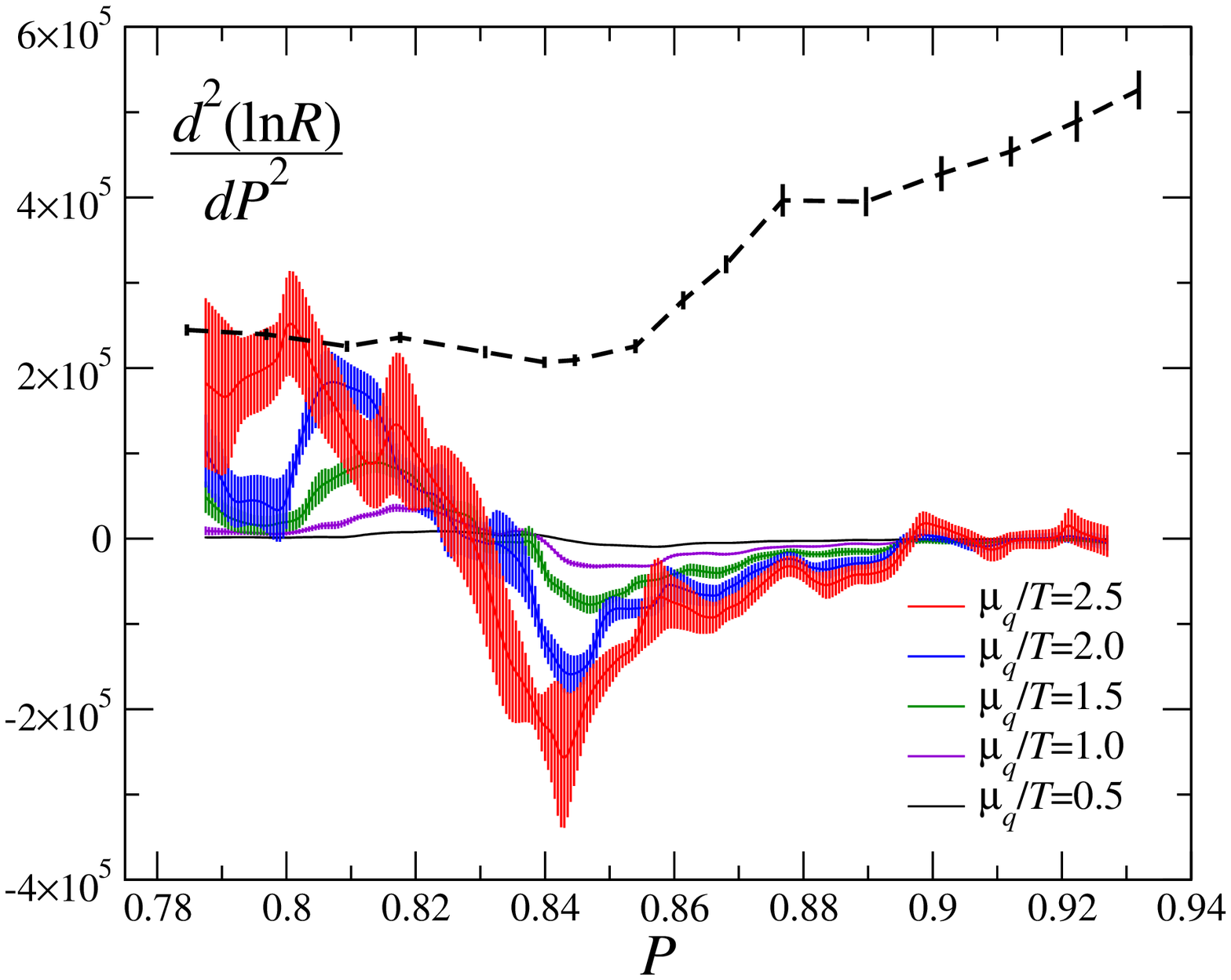}
\vskip -0.2cm
\caption{Reweigiting factor (left) and its curvature (right) as functions of $P$ \cite{eji07}.
}
\label{fig:peff}
\end{center}
\vskip -0.3cm
\end{figure}

The probability distribution function at non-zero $\mu_q$ has been 
calculated in \cite{eji07}
using the data obtained in \cite{BS05} with the 2 flavor p4-improved staggered quark action on a $16^3 \times 4$ lattice. 
The pion mass is $m_{\pi} \approx 770 {\rm MeV}$, 
which is heavier than the physical mass.
The distribution function $w(P)$ at $\mu_q=0$, i.e. the histogram of $P$, 
and the effective potential $V_{\rm eff}(P)$ are given in Fig.~\ref{fig:plaq} (right) for each $\beta$. 
The values of $\beta$ and $T/T_c$ are shown above these figures. 
The potential $V_{\rm eff}(P)$ is normalized by the minimum value 
for each $\beta$.
Because the phase transition is a crossover transition for 2 flavor QCD 
with finite quark mass, 
the distribution function is always of Gaussian type, 
i.e. the effective potential is always a quadratic function. 
The value of the plaquette at the potential minimum increases
as $\beta$ increases in accordance with the argument of 
the potential minimum.

The reweighting factor $R(P,\mu_q)$ is plotted in Fig.~\ref{fig:peff} (left).
The quark determinant $\det M(\mu_q)$ is estimated using the data of the Taylor expansion coefficients in \cite{BS05}.
Higher order terms than $\mu_q^6$ order are neglected. 
The application range was selected by checking the truncation error.
Because the sign problem is serious for the calculation of $R(P,\mu_q)$, the method  discussed in Sec.~\ref{sec:rewei} was used to avoid the sign problem.
The dashed lines in Fig.~\ref{fig:peff} (left) are the results that we obtained 
when the effect of the complex phase $e^{i \theta}$ is omitted.
Because the contribution from the complex phase is not very large, the error from the approximation to avoid the sign problem may be small.

To study the existence of a second order phase transition, 
we discuss the curvature of the potential.
The result of the curvature, $d^2 (\ln R)/dP^2 (P, \mu_q)$, is 
plotted as solid line in Fig.~\ref{fig:peff} (right). 
The magnitude of the curvature of $\ln R$ becomes larger 
as $\mu_q/T$ increases.
The dashed line in Fig.~\ref{fig:peff} (right) is the result of 
$-d^2 (\ln w)/dP^2(P)$ at $\mu_q=0$. 
This figure indicates that the maximum value of 
$d^2 (\ln R)/dP^2 (P, \mu_q)$ at $P=0.80$ becomes larger 
than $-d^2 (\ln w)/dP^2$ for $\mu_q/T \simge 2.5$. 
This means that the curvature of the effective potential 
$d^2 V_{\rm eff}/dP^2$
vanishes at $\mu_q/T \sim 2.5$ and a region of $P$ where the curvature is 
negative appears for large $\mu_q/T$, corresponding to a double-well potential. 

Further studies are, of course, necessary for the precise 
determination of the critical point in the $(T, \mu_q)$ plane, 
increasing the number of terms in the Taylor expansion of 
$\ln \det M$ and decreasing the quark mass in the simulation. 
In particular, the critical value of $\mu_q$ is sensitive to the quark mass as discussed in Sec.~\ref{sec:massdep}.
However, the argument given above suggests the appearance of 
a first order phase transition line at large $\mu_q/T$.

\subsection{Canonical approach}
\label{sec:cononi}

Another interesting approach is to construct the canonical 
partition function ${\cal Z}_{\rm C} (T,N)$ by fixing the total 
quark number $(N)$ or quark number density $(\rho)$. 
Using the canonical partition function, we can also discuss the effective potential as a function of the quark number.
In this section, we denote the grand partition function as 
${\cal Z}_{\rm GC}(T,\mu_q)$ to distinguish it from the canonical partition function 
${\cal Z}_{\rm C} (T,N)$ explicitly. 
The relation between ${\cal Z}_{\rm GC}(T,\mu_q)$ and 
${\cal Z}_{\rm C} (T,N)$ is given by 
\begin{eqnarray}
{\cal Z}_{\rm GC}(T,\mu_q) 
= \int {\cal D}U \left( \det M(\mu_q/T) \right)^{N_{\rm f}} e^{-S_g}
= \sum_{N} \ {\cal Z}_{\rm C}(T,N) e^{N \mu_q/T}. 
\label{eq:cpartition} 
\end{eqnarray}
Because this equation is a Laplace transformation from ${\cal Z}_{\rm C}$ to ${\cal Z}_{\rm GC}$ essentially, the canonical partition function is obtained from ${\cal Z}_{\rm GC} (T,\mu_q)$ by an inverse Laplace transformation.

In order to investigate the net quark number giving the largest 
contribution to ${\cal Z}_{\rm GC} (T,\mu_q)$, it is worth introducing an effective potential $V_{\rm eff}$ as 
a function of $N$, 
\begin{eqnarray}
V_{\rm eff}(N,T,\mu_q) 
\equiv - \ln {\cal Z}_{\rm C}(T,N) -N \frac{\mu_q}{T} 
= \frac{f(T,N)}{T} -N \frac{\mu_q}{T} , \hspace{3mm}
{\cal Z}_{\rm GC}(T,\mu_q) 
= \sum_{N} \ e^{-V_{\rm eff}}, 
\label{eq:effp} 
\end{eqnarray}
where $f$ is the Helmholtz free energy. 
Using the effective potential for the quark number, the argument of the nature of phase transition is possible as well as the effective potential for the plaquette, 
and the physical meaning of this effective potential is clearer than that of the plaquette. 

If there is a first order phase transition region, 
we expect that this effective potential has minima at more than one value 
of $N$. At the minima, the derivative of $V_{\rm eff}$ satisfies 
\begin{eqnarray}
\frac{\partial V_{\rm eff}}{\partial N} (N, T, \mu_q ) 
= -\frac{\partial (\ln {\cal Z}_{\rm C})}{\partial N} (T,N) 
- \frac{\mu_q}{T} =0 .
\label{eq:dpotdn} 
\end{eqnarray}
Hence, in the first order transition region of $T$, we expect 
$\partial (\ln {\cal Z}_{\rm C})/ \partial N (T,N) \equiv - \mu_q^*/T$ 
takes the same value at different $N$. 
Here, $\mu_q^* (T,N)$ is the chemical potential which gives a minimum 
of the effective potential at $(T, N)$ and becomes $\mu_q$ in the thermodynamic limit because the potential is minimized in the large volume limit.

\begin{figure}[t]
\begin{center}
\includegraphics[width=3.7in]{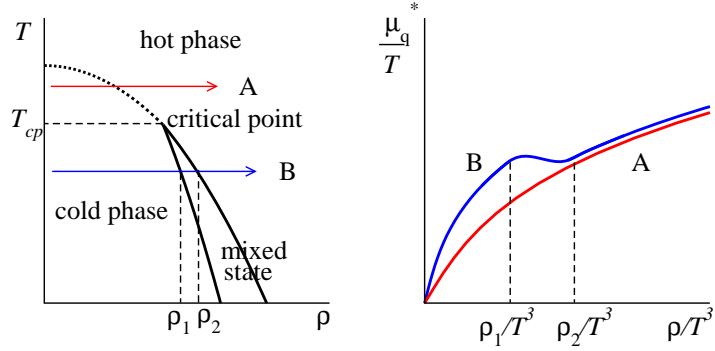}
\vskip -0.2cm
\caption{Phase structure in the $(T, \rho)$ plane and the behavior of 
$\mu_q^*/T$ as a function of $\rho$.
}
\label{fig:cano}
\end{center}
\vskip -0.3cm
\end{figure}

The phase structure in the $(T, \rho)$ plane and the expected behavior 
of $\mu_q^*/T$ are sketched in the left and right panels of 
Fig.~\ref{fig:cano}, respectively. 
The thick lines in the left figure are the phase transition line. 
We expect that the transition is crossover at low density and becomes 
of first order at high density. 
Since two states coexist on the first order transition line, the phase 
transition line splits into two lines in the high density region, 
and the two states are mixed in the region between two lines. 
The expected behavior of $\mu_q^*$ along the lines A and B are shown 
in the right figure. 
When the temperature is higher than the temperature at the critical 
point $T_{pc}$ (line A), 
$\mu_q^*$ increases monotonically as the density increases. However, 
for the case below $T_{cp}$ (line B), this line crosses the mixed state. 
Because the two states of $\rho_1$ and $\rho_2$ are realized at 
the same time, $\mu_q^*$ does not increase in this region between 
$\rho_1$ and $\rho_2$. 

Glasgow method \cite{Bar97,Gibbs86} has been a well-known method to 
compute the canonical partition function.
A few years ago, such a behavior at a first order phase transition was 
observed by Kratochvila and de Forcrand in 4 flavor QCD with staggered 
fermions \cite{Krat05} calculating 
the quark determinant by the Glasgow algorithm on a small lattice.
Also, Alexandru et al.~\cite{Alex05} proposed a method to perform simulations with canonical ensemble directly, and recently the method was tested for 2 and 4 flavor QCD \cite{Li08}.
Moreover, a method based on a saddle point approximation was proposed by \cite{eji08}. By this approximation, the computational cost is drastically reduced and the first order like behavior was observed for 2 flavor QCD. 
We explain these recent developments.

\subsubsection*{Inverse Laplace transformation}

From Eq.~(\ref{eq:cpartition}),
the canonical partition function can be obtained by an inverse Laplace 
transformation \cite{Mill87,Hase92},
\begin{eqnarray}
{\cal Z}_{\rm C}(T,N) = \frac{3}{2 \pi} \int_{-\pi/3}^{\pi/3} 
e^{-N (\mu_0/T+i\mu_i/T)} {\cal Z}_{\rm GC}(T, \mu_0+i\mu_i) \ 
d \left( \frac{\mu_i}{T} \right) ,
\label{eq:canonicalP} 
\end{eqnarray}
where $\mu_0$ is an appropriate real constant and $\mu_i$ is 
a real variable. Note that 
${\cal Z}_{\rm GC}(T, \mu_q +2\pi iT/3) = {\cal Z}_{\rm GC}(T, \mu_q)$
\cite{Rob86}.
The grand canonical partition function can be evaluated by 
the calculation of the following expectation value at $\mu_q=0$.
\begin{eqnarray}
\frac{{\cal Z}_{\rm GC}(T, \mu_q)}{{\cal Z}_{\rm GC}(T,0)}
&=& \frac{1}{{\cal Z}_{\rm GC}} \int {\cal D}U 
\left( \frac{\det M(\mu_q/T)}{\det M(0)} \right)^{N_{\rm f}}
\det M(0) ^{N_{\rm f}} e^{-S_g} 
= \left\langle 
\left( \frac{\det M(\mu_q/T)}{\det M(0)} \right)^{N_{\rm f}}
\right\rangle_{(T, \mu_q=0)} . 
\nonumber \\ &&
\label{eq:normZGC} 
\end{eqnarray}

If one adopts $\mu_0=0$, the chemical potential in Eq.~(\ref{eq:normZGC}) is purely imaginary and the quark determinant is real. Therefore, the calculation of the quark determinant at $i \mu_i$ is easier than the calculation at real $\mu_q$.
Performing a simulation at $\mu_q=0$, the standard staggered quark determinant can be computed for any chemical potential at the cost of diagonalizing a matrix modified from the staggered fermion matrix by using the Glasgow algorithm \cite{Gibbs86}. 
Because Eq.~(\ref{eq:canonicalP}) is actually a Fourier transformation, if one calculates the Fourier coefficients of $\det M(i\mu_i/T)$ for each configuration, 
the canonical partition function ${\cal Z}_{\rm C}(T, N) / {\cal Z}_{\rm GC}(T,0)$ can be computed by averaging them over the configurations.

Kratochvila and de Forcrand calculated $-\partial (\ln {\cal Z}_{\rm C})/ \partial N (T,N)$ as a function of the baryon number $N_B \equiv N/3$ with a fixed temperature $(\beta)$ performing a simulation of 4 flavor QCD with staggered quarks on a $6^3 \times 4$ lattice \cite{Krat05}. 
The phase transition is of first order even at $\mu_q=0$ for their simulation of 4 flavor QCD. 
Their result of $-\partial (\ln {\cal Z}_{\rm C})/ \partial N (T,N)$ shows an S-shape in the temperature range below $T_c$, as we discussed above for a first order phase transition.

\subsubsection*{Simulations with canonical ensemble}

\begin{figure}[t]
\begin{center}
\includegraphics[width=2.8in]{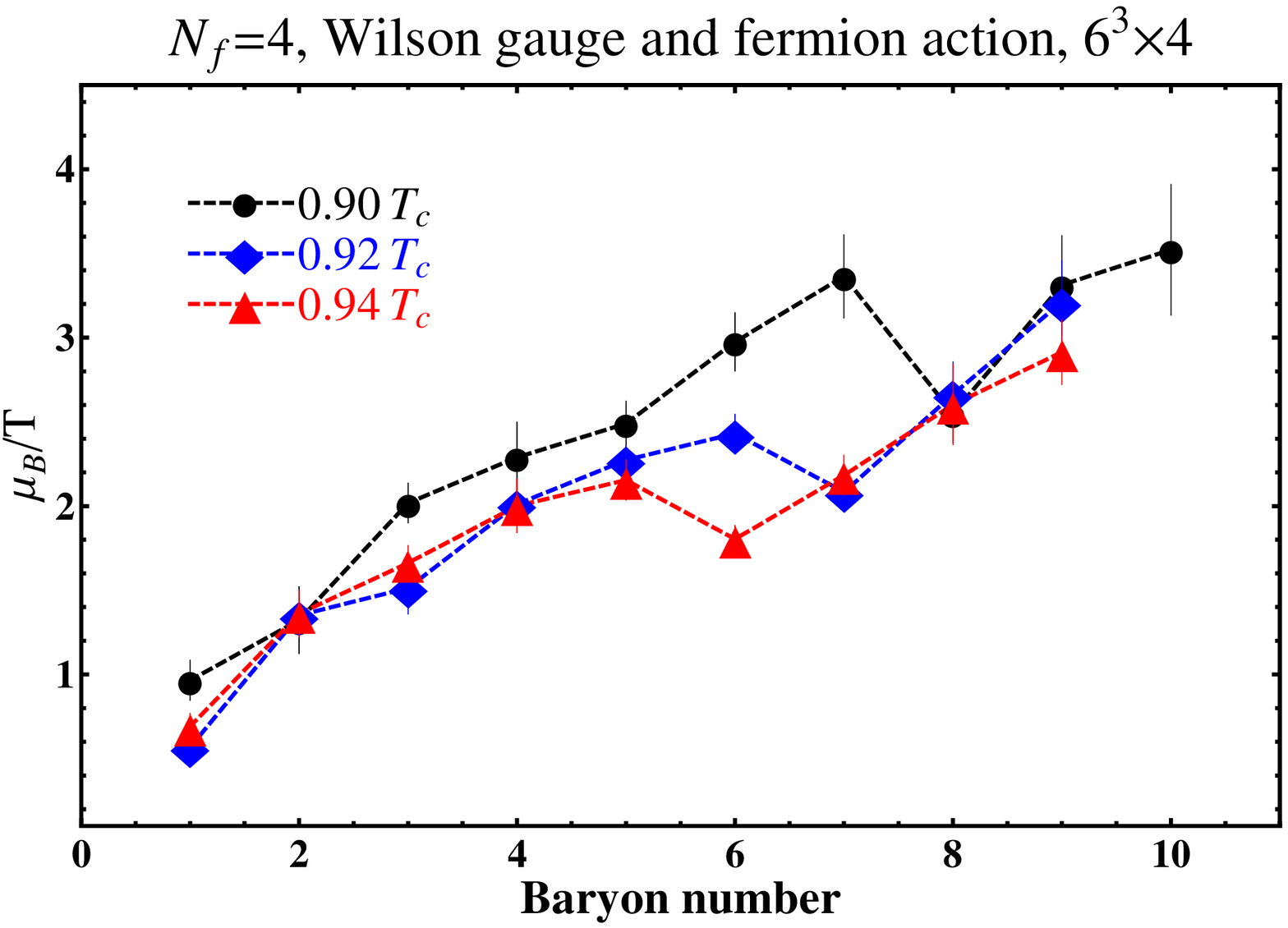}
\hskip 0.2cm
\includegraphics[width=2.8in]{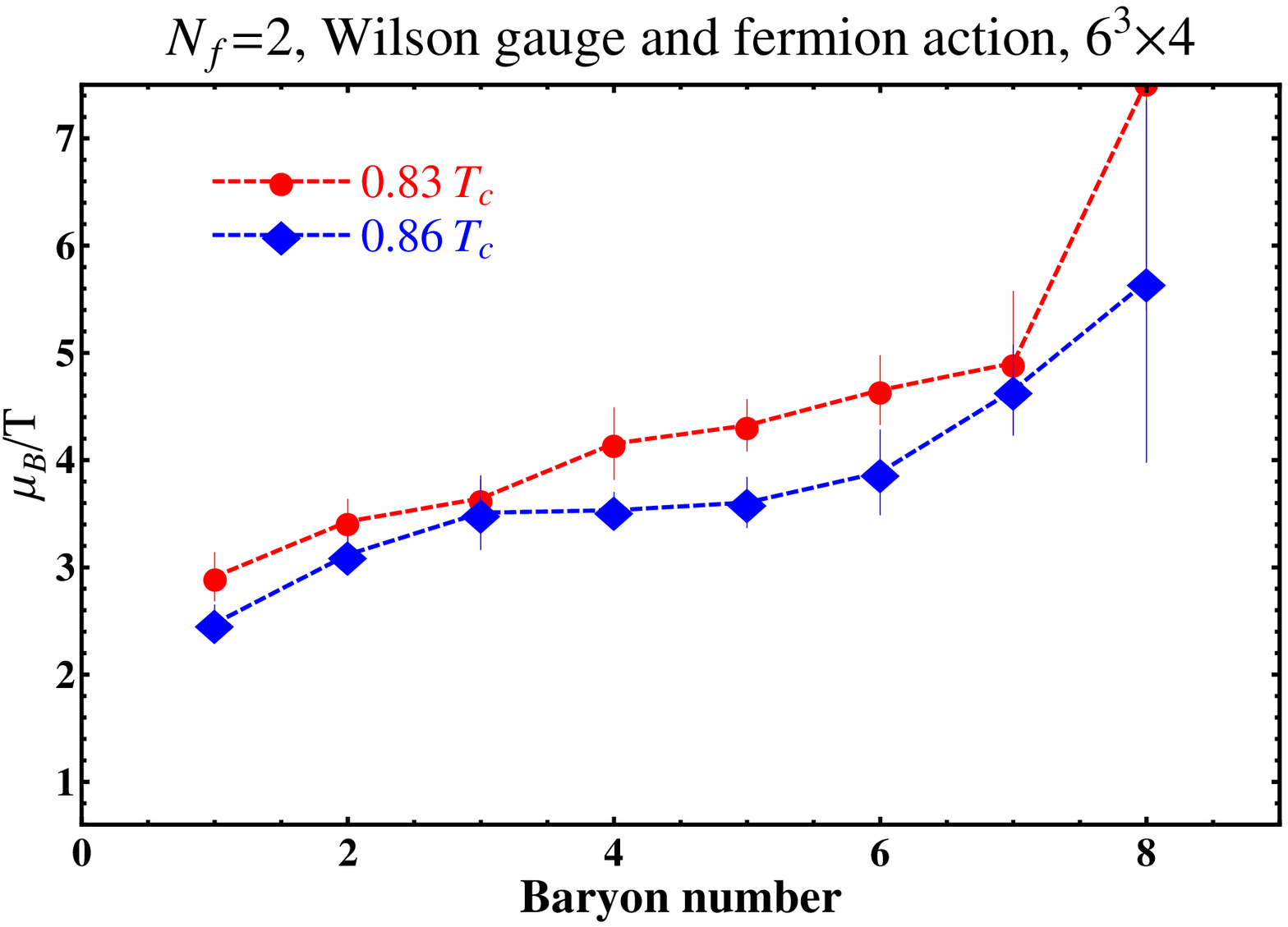}
\vskip -0.2cm
\caption{Baryon chemical potential as a function of baryon number for $N_f=4$ (left) 
$N_f=2$ (right) obtained by simulations with canonical ensemble \cite{Li08}.
}
\label{fig:kent}
\end{center}
\vskip -0.3cm
\end{figure} 

By replacing the order of the integrals of the inverse Laplace 
transformation Eq.~(\ref{eq:canonicalP}) and the path integral 
in ${\cal Z_{\rm GC}}$, the canonical partition function is written by 
\begin{eqnarray}
{\cal Z}_{\rm C}(T,N) = \int {\cal D}U e^{-S_g}
{\det}_N M^{N_{\rm f}}, 
\label{eq:canoDS} 
\end{eqnarray}
where $\det_N M^{N_{\rm f}}$ is defined by
\begin{eqnarray}
{\det}_N M^{N_{\rm f}} = \frac{1}{2 \pi} \int_{0}^{2\pi} 
e^{-iN \mu_i/T} \det M(i\mu_i/T)^{N_{\rm f}} \ d \left( \frac{\mu_i}{T} \right) .
\label{eq:detMfour} 
\end{eqnarray}
The method to perform Monte-Carlo simulations with this partition function has been proposed by \cite{Alex05,Meng08}. 
Although ${\det}_N M^{N_{\rm f}}$ is complex, 
we obtain ${\det}_N M^{N_{\rm f}} = \left( {\det}_{-N} M^{N_{\rm f}} \right)^{*}$ 
and ${\cal Z}_{\rm C}(T,N)= {\cal Z}_{\rm C}(T,-N)$
from a symmetry under the replacement from $\mu_q$ to $-\mu_q$. 
Using these properties, we can rewrite the partition function as
\begin{eqnarray}
{\cal Z}_{\rm C}(T,N) = \int {\cal D}U e^{-S_g}
{\rm Re} \left( {\det}_N M^{N_{\rm f}} \right). 
\label{eq:canoDSRe} 
\end{eqnarray}
Then, the Boltzmann weight is real and the Monte-Carlo method is applicable. 
One can generate configurations according to the weight 
$ \exp(-S_g) {\rm Re} (\det_N M^{N_{\rm f}} )$, and 
$\det_N M^{N_{\rm f}}$ is computed numerically with an approximation, 
\begin{eqnarray}
{\det}_N M^{N_{\rm f}} \approx \frac{1}{N_{\rm dis}} \sum_{j=0}^{N_{\rm dis}-1} 
e^{-2 \pi ijN /N_{\rm dis}} \det M(2 \pi ij/N_{\rm dis})^{N_{\rm f}} .
\label{eq:detMfourap} 
\end{eqnarray}
This approximation is applicable for $N_{\rm dis} \gg N$. 

The $\chi$QCD collaboration (Kentucky group) performed simulations with the canonical ensemble on a $6^3 \times 4$ lattice for 2 and 4 flavor QCD with 
Wilson quarks using this method and the preliminary results are presented 
in this conference \cite{Li08}. 
The results of the chemical potential 
$\mu_B^*/T \equiv -\partial( \ln {\cal Z}_{\rm C})/ \partial (N/3)$ 
are shown in Fig.~\ref{fig:kent} for 4 flavor QCD (left) and 2 flavor QCD (right). 
Their result of 4 flavor QCD shows an S-shape in the temperature range below $T_c$, suggesting a first order phase transition. This result is consistent with the previous result \cite{Krat05}. On the other hand, the chemical potential increases monotonically for 2 flavor QCD. This result suggests that the phase transition is crossover at the temperature they investigated.

\subsubsection*{Saddle point approximation}

\begin{figure}[t]
\begin{center}
\includegraphics[width=2.7in]{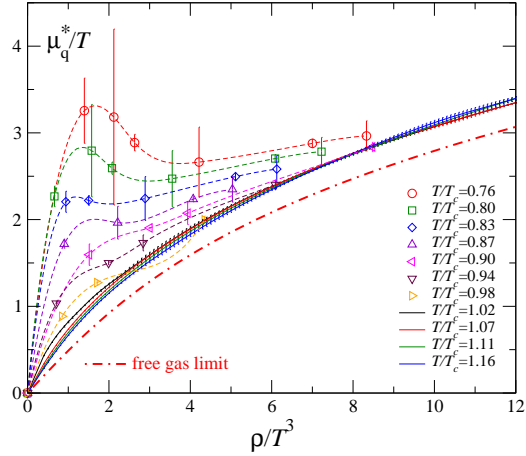}
\vskip -0.2cm
\caption{Chemical potential vs. quark number density for $N_f=2$ with a saddle point approximation \cite{eji08}.
}
\label{fig:chem}
\end{center}
\vskip -0.3cm
\end{figure} 

However, the studies by above-mentioned two methods need much computational cost and are difficult except on a small lattice with present day computer resources.
To reduce the computational cost, a method based on a saddle point approximation has been proposed in \cite{eji08}. 
If one selects a saddle point as $\mu_0$ in Eq.~(\ref{eq:canonicalP}). The information which is needed for the integral is only around the saddle point when the volume is sufficiently large. 
Moreover, if we restrict ourselves to study the low density region, the value of $\det M (\mu_q/T)$ near the saddle point can be estimated by the Taylor expansion around $\mu_q=0$. 
The calculations by the Taylor expansion are much cheaper than the exact calculations and the studies using large lattices are possible. 
Also, the truncation error can be systematically controlled by increasing the number of the expansion coefficients.

We perform the integral in Eq.~(\ref{eq:canonicalP}) by 
a saddle point approximation.
We denote the quark number density in a lattice unit and physical unit as
$\bar{\rho}=N/N_s^3$ and $\rho/T^3=\bar{\rho} N_t^3$, respectively. 
We assume that a saddle point $z_0$ exists in the complex $\mu_q/T$ 
plane for each configuration, which satisfies
$D'(z_0) - \bar{\rho} =0$, 
where $(\det M(z)/\det M(0))^{N_{\rm f}} = \exp[N_s^3 D(z)] $ and 
$D'(z)=dD(z)/dz$.
We then perform a Taylor expansion around the saddle point and obtain 
the canonical partition function, 
\begin{eqnarray}
{\cal Z}_{\rm C}(T, \bar{\rho} V) 
&=& \frac{3}{2 \pi} {\cal Z}_{\rm GC}(T,0) \left\langle \int_{-\pi/3}^{\pi/3} 
\exp \left[ V \left(D (z_0) - \bar{\rho} z_0 
- \frac{1}{2} D'' (z_0) x^2 + \cdots \right) \right] 
dx \right\rangle_{(T, \mu_q=0)} \nonumber \\
& \approx & \frac{3}{\sqrt{2 \pi}} {\cal Z}_{\rm GC} (T,0)
\left\langle \exp \left[ V \left( D(z_0) - \bar{\rho} z_0 \right) \right] 
e^{-i \alpha/2} \sqrt{ \frac{1}{V |D''(z_0)|}}
\right\rangle_{(T, \mu_q=0)} .
\label{eq:zcspa}
\end{eqnarray}
Here, $ D''(z) = d^2 D(z) /dz^2,$ $V \equiv N_s^3$ and 
$D''(z)=|D''(z)| e^{i \alpha}$.
We chose a path which passes the saddle point.
Higher order terms in the expansion of $D(z)$ become negligible when 
the volume $V$ is sufficiently large. 

We calculate the derivative of the effective potential with respect to $N$ or $\rho$.
Within the framework of the saddle point approximation, 
this quantity can be evaluated by 
\begin{eqnarray}
\frac{\mu_q^*}{T} 
= - \frac{1}{V} \frac{\partial \ln {\cal Z}_C (T, \bar{\rho} V)}
{\partial \bar{\rho}}
\approx \frac{
\left\langle z_0 \ \exp \left[ V \left( D(z_0)
- \bar{\rho} z_0 \right) \right] 
e^{-i \alpha /2} \sqrt{ \frac{1}{V |D''(z_0)|}}
\right\rangle_{(T, \mu_q=0)}}{
\left\langle \exp \left[ V \left( D(z_0) 
- \bar{\rho} z_0 \right) \right] 
e^{-i \alpha /2} \sqrt{ \frac{1}{V |D''(z_0)|}}
\right\rangle_{(T, \mu_q=0)}}. 
\label{eq:chemap}
\end{eqnarray}
This equation is similar to the formula of the reweighting method 
for finite $\mu_q$. 
The operator in the denominator corresponds to a reweighting factor, 
and $\mu_q^* /T$ is an expectation value of the saddle 
point calculated with this modification factor.

The derivative of $\ln {\cal Z}_C$ was computed in \cite{eji08} using the data obtained in \cite{BS05} with the 2 flavor p4-improved staggered quark action, $m_{\pi} \approx 770 {\rm MeV}$. 
Because the modification factor is a complex number, this calculation suffers from the sign problem. To eliminate the sign problem, the approximation discussed in Sec.~\ref{sec:rewei} was used. If one assumes that the distribution of the complex phase is well-approximated by a Gaussian function, the complex phase factor $e^{i \theta}$ can be replaced by $\exp[- \langle \theta^2 \rangle /2]$. 
Moreover, the quark determinant was estimated by the Taylor expansion up to $O(\mu_q^6)$. 
Because the calculation of Eq.~(\ref{eq:chemap}) is similar to the calculation by the reweighting method, the configurations which give important contribution are changed by the modification factor. 
To avoid this problem, the multi-$\beta$ reweighting method \cite{Swen89} was used. 
By this method, the important configurations are automatically selected among all configurations generated at many simulation points of $\beta$.

The result of $\mu_q^*/T$ is shown in Fig.~\ref{fig:chem} as a function of $\rho/T^3$ for each temperature $T/T_c (\beta)$. 
The dot-dashed line is the value of the free quark-gluon gas in 
the continuum theory, 
$\rho/T^3 = N_{\rm f} [ (\mu_q/T) + (1/\pi^2) (\mu_q/T)^3]$.
From this figure, we find that a qualitative feature of $\mu_q^*/T$ 
changes around $T/T_c \sim 0.8$, i.e. $\mu_q^*/T$ increases monotonically 
as $\rho$ increases above 0.8, whereas it shows an S-shape below 0.8. 
This means that there is more than one value of $\rho/T^3$ for 
one value of $\mu_q^*/T$ below $T/T_c \sim 0.8$.
This is a signature of a first order phase transition. 
Although some approximations are used, the critical value of $\mu_q^*/T$ is about $2.4$, which is roughly consistent with the critical point estimated in Sec.~\ref{sec:poten} by calculating the effective potential of the plaquette using the same configurations, $(T/T_c, \mu_q/T) \approx (0.76, 2.5)$. 
The difference between these two results may be a systematic error. 
Further studies are necessary to predict the critical point quantitatively, but we find that the canonical approach is useful to study the phase structure at finite density.

\section{Summary}
\label{sec:summary} 

We reviewed recent studies of lattice QCD at finite density.
Remarkable progress in the study of the equation of state was obtained. The MILC Collaboration and the RBC-Bielefeld Collaboration performed simulations with quark masses near the physical point using improved staggered quark actions and calculated thermodynamic quantities along lines of constant entropy per baryon number in the low density region using the Taylor expansion method. Because experimental results obtained in heavy-ion collisions can be well-explained by a perfect fluid model, the isentropic equation of state is needed for the analysis of the experimental data. 
Moreover, fluctuations of hadron numbers at finite density are important, which can be measured in the experiments. If there is a critical point, the fluctuation of baryon should be large around that point.
Recent simulations show that the baryon number fluctuation makes a peak near $T_c$ at finite $\mu_q$ and the fluctuation becomes larger when the quark masses are decreased to the physical point.
Confirmations by other quark actions are also important.  
Simulations with a Wilson type quark action were performed by the WHOT-QCD Collaboration and the large fluctuation at finite $\mu_q$ was discussed. 

The chemical potential dependence of the critical line in 2+1 flavor QCD with quark masses $(m_{ud}, m_s)$ was studied at low density to understand the grovel structure of the critical surface in the $(m_{ud}, m_s, \mu_q)$ parameter space.
The $\mu_q$ dependence of the critical quark mass is found to be small in the low density regime and the current result of $\partial m_c/ \partial (\mu_q^2)$ is slightly negative at $m_{ud}=m_s$.
Further studies in a wide range of the parameter space are important to understand the phase structure.

Some methods to investigate finite density QCD beyond the low density region were also discussed. 
A method based on the investigation of an effective potential as a function of the average plaquette was proposed introducing an approximation to avoid the sign problem, and the existence of the critical point at finite density was suggested by a simulation with improved staggered quarks.
Moreover, it was found that interesting information about the order of phase transitions at finite density is obtained by constructing the canonical partition function for each quark number.

\section*{Acknowledgements}
I wish to thank F. Karsch, K. Kanaya, P. de Forcrand, C. DeTar, M.-P. Lombardo,
S. J. Hands, O. Philipsen, K. Splittorff, L. Levkova, K. Fukushima, A. Li, 
C. Miao, C. Schmidt for helpful discussions, comments, and preparing figures. 
This work has been authored under Contract No.~DE-AC02-98CH10886
with the U.S. Department of Energy.

\end{document}